\documentclass[12pt]{article}

\usepackage{url,hyperref,microtype,booktabs,amssymb,amsmath,multirow,array,graphicx,xcolor}
\usepackage[margin=2cm]{geometry}
\usepackage{authblk}
\usepackage[backend=bibtex,style=numeric]{biblatex}
\usepackage[onehalfspacing]{setspace}
\newcommand{\gray}[1]{\textcolor{gray}{#1}}

\graphicspath{{figures/}}

\title{Visualizing and Interpreting Unsupervised Solar Wind Classifications} 

\author[1]{\small Jorge Amaya}
\author[1]{Romain Dupuis}
\author[1]{Maria Elena Innocenti}
\author[1]{Giovanni Lapenta}
\affil[1]{Centre for mathematical Plasma-Astrophysics, CmPA, Mathematics Department, KU Leuven, University of Leuven, Belgium}
\date{}

\begin{document}
	
\onecolumn

\maketitle

{\tiny \textbf{Word count}: in text (10172), in headers (105), outside text (843). Number of floats/tables/figures: 14.}

\begin{abstract}

One of the goals of machine learning is to eliminate tedious and arduous repetitive work. The manual and semi-automatic classification of millions of hours of solar wind data from multiple missions can be replaced by automatic algorithms that can discover, in mountains of multi-dimensional data, the real differences in the solar wind properties. In this paper we present how unsupervised clustering techniques can be used to segregate different types of solar wind. We propose the use of advanced data reduction methods to pre-process the data, and we introduce the use of Self-Organizing Maps to visualize and interpret 14 years of ACE data. Finally, we show how these techniques can potentially be used to uncover hidden information, and how they compare with previous manual and automatic categorizations.

{\small \textbf{Keywords:} solar wind, ACE, Self-Organizing Maps, clustering, autoencoder, PCA, unsupervised, machine learning}
\end{abstract}
	
	\section{Introduction}
	The effects of solar activity on the magnetic environment of the Earth have been observed since the publication of Edward Sabine's work in 1852 \cite{Sabine1852}. During almost two hundred years we have learned about the intimate connection between our star and the plasma environment of the Earth. Three main physical processes connect us to the Sun: the transfer of electromagnetic radiation, the transport of energetic particles, and the flow of solar wind. The later is a continuous stream of charged particles that carries the solar magnetic field out of the corona and into the interplanetary space.
	
	The name \textit{solar wind} was coined by Parker in 1958 because `the gross dynamical properties of the outward streaming gas [from the Sun] are hydrodynamic in character'\cite{Parker1958}. Over time we have learned that the wind also has many more complex properties. Initially, it was natural to classify the solar wind by defining a boundary between \textit{fast} and \textit{slow} winds \cite{Neugebauer1966,Schwenn1983,Schwenn1990,Habbal1997}. The former has been associated with mean speed values of 750 km/s (or in some publications with values larger than 600 km/s), while the later shows a limit at 500 km/s, where the compositional ratio (Fe/O) shows a break \cite{Feldman2005,Stakhiv2015}. The solar wind also carries information about its origins on the Sun. At certain solar distances the ion composition of the solar wind is expected to be frozen-in, reflecting the electron temperature in the corona and its region of origin \cite{Feldman2005,Zhao2009,Stakhiv2015}. These particles have multiple energies and show a variety of kinetic properties, including non-Maxwellian velocity distributions \cite{Pierrard2010,Matteini2012}.
	
	The solar wind is also connected to the Sun by Interplanetary Magnetic Field (IMF) lines directed towards the Sun, away from the Sun, or in the case of flux ropes, connected at both ends \cite{Owens2016,Gosling2010}. The region separating IMF lines of opposite polarity (directed away or towards the Sun) is called the Heliospherc Current Sheet (HCS) \cite{Smith2001}. When a spacecraft crosses the HCS instruments onboard measure the change in polarity of the magnetic field. In quiet wind conditions the plasma around the HCS presents discontinuities in density, temperature velocity and magnetic field \cite{Eselevich1988}. This perturbed region surrounding the HCS is called the Heliospheric Plasma Sheet (HPS). The passage of the spacecraft from one side of the HPS to the other is known as a Sector Boundary Crossing (SBC) \cite{Winterhalter1994}. In spacecraft observations these are sometimes confused with Corotating Interaction Regions (CIR), which are zones of the solar wind where fast flows have caught up with slow downstream solar wind, compressing the plasma \cite{1980ApJ...237..620F,Richardson2004}.
	
	From the point of view of a spacecraft SBCs and CIRs can show similar sudden changes in the plasma properties. These two in turn are often grouped and mixed with other transient events, like Coronal Mass Ejections (CME) and Magnetic Clouds (MC). Since 1981 when \cite{Burlaga1981} described the propagation of MC behind an interplanetary shock, it was suspected that CMEs and MC where coupled. However, more recent studies show that CMEs observed near the Sun do not necessarily become MC, but instead `pressure pulses' \cite{Gopalswamy1998,Wu2006}.
	
	Much more recently it has been revealed, by observations from Parker Solar Probe, that the properties of the solar wind can be drastically different closer to the Sun, were the plasma flow is more pristine and has not yet mixed with the interplanetary environment. Patches of large intermittent magnetic field reversals, associated with jets of plasma and enhanced Poynting flux, have been observed and named `switchbacks' \cite{Bale2019,Bandyopadhyay2020}.
	
	The solar wind is thus not only an hydrodynamic flow, but a compressible mix of different populations of charged particles and electromagnetic fields that carry information of their solar origin (helmet streamer, coronal holes, filaments, solar active regions, etc.) and is the dominion of complex plasma interactions (ICMEs, MC, CIRs, SBCs, switchbacks).
	
	To identify and study each one of these phenomena we have relied in the past on a manual search, identification and classification of spacecraft data. Multiple authors have created empirical methods of wind type identification based on in-situ satellite observations and remote imaging of the solar corona. Over the years the number and types of solar wind classes has changed, following our understanding of the complexity of heliospheric physics.
	
	Solar wind classification serves four main roles:
	\begin{enumerate}
		\item it is used for the characterization of its origins in the corona,
		\item to identify the conditions where the solar wind is geoeffective,
		\item to isolate different plasma populations in order to perform statistical analysis,
		\item to study the basic transport effects of space plasmas of different nature.
	\end{enumerate}
	
	Among the existing classifications we can include the original review work by \cite{Withbroe1986}, the impressive continuous inventory by \cite{Richardson2000,Richardson2010,Richardson2012}, and the detailed studies by \cite{Zhao2009} and \cite{Xu2015b}. These publications classify the solar wind based on their ion composition,  and on the transient events detected. Each system includes two, three or four classes, generally involving coronal-hole origins, CMEs, streamer belt origins and sector reversal regions.
	
	The precise point of origin of the solar wind can be traced back from spacecraft positions to the solar corona and the photosphere: multiple authors \cite{Neugebauer2002,Zhao2009,Fu2015,Zhao2017} have used a ballistic approximation coupled to a Potential Field Source Surface (PFSS) model to trace back solar wind observations to their original sources on the Sun. This procedure relies on multiple assumptions, including a constant solar wind speed and a force free magnetic field configuration of the solar corona. The uncertainty on the source position is estimated around $\pm 10^{\circ}$ by  \cite{Neugebauer2002}. This is currently the best method to acquire the \textit{ground truth} about the origin of the solar wind. Unfortunately, to our knowledge, there is no central repository of solar wind origins for any space mission that we can use to train or verify our novel machine learning techniques.
	
	We are moving now towards a new era of data analysis, where manual human intervention can be replaced by \textit{intelligent} software. The trend has already started, with the work by \cite{Camporeale2017b} who used the \cite{Xu2015b} classes to train a Gaussian Process algorithm that autonomously assigns the solar wind to the proper class, and by \cite{Roberts2020} who used unsupervised classification to perform a 4 and 8 class solar wind classification. A recent publication by \cite{Bloch2020} uses unsupervised techniques to classify ACE and Ulysses observations, and \cite{Li2020} have successfully tested ten different supervised techniques to reproduce the categories introduced by \cite{Xu2015b}.

	The most basic ML techniques learn using two approaches: a) in supervised learning the algorithms are shown a group of inputs, $\mathbf{X} \in\mathbb{R}^n$, and outputs, $\mathbf{Y}\in\mathbb{R}^o$, with the goal of finding a non-linear relationship between them, $\xi_s: \mathbf{X} \rightarrow \mathbf{Y}$, b) in unsupervised learning the machine is presented with a cloud of multi-dimensional points, $\mathbf{X}\in\mathbb{R}^n$, that have to be autonomously categorized in different classes, either performing associations with representative points in the same data space, $\xi_u: \mathbf{X} \rightarrow \mathbf{W}\in\mathbb{R}^n$, or by grouping neighboring data points together into an assigned set, $\xi_u: \mathbf{X} \rightarrow g\in\mathbb{R}$. This means that we can program the computer to learn about the different types of solar wind using the existing empirical classifications using method (a), or allowing the computer to independently detect patterns in the solar wind properties with method (b).
	
	In the present work we show how the second method, unsupervised classification, can be used to segregate different types of solar wind. In addition, we show how to visualize and interpret such results. The goal of this paper is to introduce the use of unsupervised techniques to our community, including the best use practices and the opportunities that such methods can bring. We promote the use of one specific type of classification, called Self-Organizing Maps, and we compare it to simpler classification techniques.
	
	In the next sections we present in detail the techniques of data processing (section \ref{sec:dataprocessing}), data dimension reduction (sections \ref{sec:reducpca}, \ref{sec:reduckpca} and \ref{sec:reducae}) and data clustering (section \ref{sec:clustering}) that we have used. We then present in detail the Self-Organizing Map technique and all its properties in section \ref{sec:som}. We show how to connect all of these parts together in section \ref{sec:fullarchi}, and finally we show how the full system can be used to study 14 years of solar wind data from the ACE spacecraft in section \ref{sec:results}.
	
	\section{Materials and Methods}
	
	\subsection{Data and Processing}
	\label{sec:dataprocessing}
	\subsubsection{Data Set Used}
	The solar wind data used in this work was obtained by the Advanced Composition Explorer (ACE) spacecraft, during a period of 14 years, between 1998 and 2011. The data can be downloaded from the \href{ftp://mussel.srl.caltech.edu/pub/ace/level2/multi}{FTP servers of The ACE Science Center (ASC)} \cite{Garrard1998}. The files in this repository correspond to a compilation of hourly average data from four instruments: MAG (Magnetometer) \cite{Smith1998}, SWEPAM (Solar Wind Electron, Proton, and Alpha Monitor) \cite{McComas1998}, EPAM (Electron, Proton, and Alpha Monitor) \cite{Gold1998}, and SWICS (Solar Wind Ion Composition Spectrometer) \cite{Gloeckler1998}. A detailed description of the entries in this data set can be found in the \href{http://www.srl.caltech.edu/ACE/ASC/level2/lvl2DATA_MULTI.html}{ASC website} listed in section \ref{sec:repos}.
	
	A total of 122712 data points are available. However, routine maintenance operations, low statistics, instrument saturation and degradation produce gaps and errors in the data. The SWICS data includes a flag assessing the quality of the calculated plasma moments. We retain only \textit{Good quality} entries. Our pre-processed data set contains a total of 72454 points.
	
	\subsubsection{Additional Derived Features}
	We created additional features for each entry, based on previous knowledge of the physical properties of the solar wind. Some are derived from the existing properties in the data set, others computed from statistical analysis of their evolution. We introduce here the additional \textit{engineered} features included in our data set.
	
	Multiple techniques have been proposed in the literature to identify ejecta, Interplanetary Coronal Mass Ejections (ICME), and solar wind origins in the ACE data. \cite{Zhao2009} suggest that, during solar cycle 23, three classes of solar wind can be identified using its speed, $V_{sw}$, and the oxygen ion charge state ratio, $O^{7+}/O^{6+}$. It has been shown that slow winds originating in coronal streamers correlate with high values of the charge state ratio and fast winds coming from coronal holes present low values \cite{Schwenn1983,Withbroe1986,Schwenn1990}. Plasma formed in coronal loops associated with CMEs also show high values of the charge state ratio \cite{Xu2015b,Zhao2017}. The classification boundaries of the Z09 model, proposed by \cite{Zhao2009}, are presented in Table \ref{tab:swtypes}.
	
	\cite{Xu2015b} suggested an alternative four classes system based on the proton-specific entropy, $S_p = T_p/n_p^{2/3}$ [K cm$^2$], the Alfv\'en speed, $V_A = B / (\mu_0 m_p n_p)^{1/2}$ [Km s$^{-1}$], and the ratio between the expected and the measured proton temperature, $T_\text{exp}/T_p = (V_{sw}/258)^{3.113}/T_p$ [-], where $n_p$ is the proton number density, $m_p$ is the proton mass, and $\mu_0$ is the permeability of free space. The classification boundaries used for the X15 model, proposed by \cite{Xu2015b}, are also presented in Table \ref{tab:swtypes}. For each entry in the data set we have included the values of $S_p$, $V_A$, $T_\text{exp}$, $T_\text{ratio}=T_\text{exp}/T_p$, and the solar wind type.
	
	Two additional empirical threshold methods will be included in this work for comparison. These two methods were derived from the compositional observations of the solar wind at higher heliospheric latitudes, using data from the \textit{Ulysses} mission \cite{Wenzel1992}. The first model, that we call vS15, comes from the work by \cite{VonSteiger2015}, where the first figure shows a clear division between Coronal Hole (CH) sources and non-Coronal Hole (NCH) wind. The boundary between the two classes is presented in Table \ref{tab:swtypes}. The second threshold model was presented as an example by \cite{Bloch2020}. This boundary, named here B20, is an empirical approximation that divides CH and NCH origin winds. The threshold values are shown in Table \ref{tab:swtypes}.
	
	In addition to the instantaneous properties of the solar wind used in all previous classifications, we can perform statistical operations over a window of time of six hours, including values of the maximum, minimum, mean, standard deviation, variance, auto-correlation, and range. We expect to capture with some of these quantities turbulent signals or sudden jumps associated with different transient events. These additional rolling operations are a complement to the stationary solar wind parameters mentioned above and add information about the temporal evolution of the plasma. The selection of the statistical parameters and the window of time is arbitrary and will require a closer examination in the future.

	An additional term, which has been successfully used in the study of solar wind turbulence \cite{Zhao2018,Adhikari2020,Magyar2019,DAmicis2015}, is included here to account for additional time correlations. The normalized cross-helicity, $\sigma_c$, is defined in eq. \eqref{eq:sigmac}, where $\boldsymbol{b} = \left(\boldsymbol{B}- \boldsymbol{\left<B\right>}\right)/(\mu_0m_pn_p)^{1/2}$ is the fluctuating magnetic field in Alfv\'en units, $\boldsymbol{v} = \boldsymbol{V_{sw}}- \boldsymbol{\left<V_{sw}\right>}$ is the fluctuating solar wind velocity, and $\left<.\right>$ denotes the averaging of quantities over a time window of three hours \cite{Roberts2020}.
	
	\begin{align}
	\sigma_c & = 2 \left< \boldsymbol{b}\cdot\boldsymbol{v}\right>/\left<\boldsymbol{b}^2 + \boldsymbol{v}^2\right> \label{eq:sigmac} %
	\end{align}
	
	Due to gaps in the data, some of the above quantities can not be obtained. We eliminate from the data set all entries for which the derived features presented in this section could not be calculated. This leaves a total of 51374 entries in the data set used in the present work.
	
	To account for the differences in units and scale, each feature column $\boldsymbol{F}$ in the data set is normalized to values between 0 and 1, using: $\boldsymbol{f}=\left(\boldsymbol{F}-\min{\boldsymbol{F}}\right) /\left(\max{\boldsymbol{F}}-\min{\boldsymbol{F}}\right)$.
	
	Not all the features might be useful and some of them can be strongly correlated. We do not perform here a detailed evaluation of the inter dependencies of the different features, and we leave that task for a future work. The present manuscript focuses on the description of the methodology and on the visualization and interpretation capabilities of unsupervised machine learning classification. We limit our work here to test and compare a single model that incorporates a total of 15 features. These are listed in table \ref{tab:features}. 
	
	\subsubsection{Complementary Data Catalogs}
	\label{sec:catalogs}
	We support the interpretation of our results using data from three solar wind event catalogs. The first is the well known Cane and Richardson catalog that contains information about ICMEs detected in the solar wind in front of the Earth \cite{Cane2003} \cite{Richardson2010} \footnote{\href{http://www.srl.caltech.edu/ACE/ASC/DATA/level3/icmetable2.htm}{Near-Earth Interplanetary Coronal Mass Ejections Since January 1996: http://www.srl.caltech.edu/ACE/ASC/DATA/level3/icmetable2.htm}}. We used the August 16, 2019 revision. As the authors state in their website, there is no spreadsheet or text version of this catalog and offline editing was necessary. We downloaded and re-formatted the catalog to use it in our application. The CSV file created has been made available in our repository. We call this, the Richardson and Cane catalog. 
	
	The second catalog corresponds to the ACE List of Disturbances and Transients\footnote{\href{http://www.ssg.sr.unh.edu/mag/ace/ACElists/obs\_list.html}{ACE Lists of Disturbances and Transients: http://www.ssg.sr.unh.edu/mag/ace/ACElists/obs\_list.html}} produced by the University of New Hampshire. As in the previous case, the catalog is only available as an html webpage, so we have manually edited the file and extracted the catalog data into a file also available in our repository. This is hereafter referred to as the UNH catalog.
	
	Finally, we also included data from the Shock Database\footnote{\href{https://www.cfa.harvard.edu/shocks/ac_master_data/}{Harvard-Smithsonian, Center for Astrophysics, Interplanetary Shock Database - ACE: https://www.cfa.harvard.edu/shocks/ac\_master\_data/}} maintained by Dr. Michael L. Stevens and Professor Justin C. Kasper at the Harvard-Smithsonian Center for Astrophysics. Once again we have gathered and edited multiple web-pages in a single file available in our repository. In this work this database will be known as the CfA catalog.
	
	\subsection{Dimension Reduction and Clustering}
	\subsubsection{Dimension Reduction using PCA}
	\label{sec:reducpca}
	Principal Component Analysis (PCA) is a mathematical tool used in data analysis to simplify and extract the most relevant features in a complex data set. This technique is used to create entries composed of linearly independent \textit{principal components}. These are the eigenvectors, $\boldsymbol v$, of the covariance matrix $\boldsymbol\Sigma = (\Sigma_{ij})$ applied to the centered data, eq.\eqref{eq:covariance}, ordered from the largest to the smallest eigenvalue, $\lambda_1 \ge \lambda_2 \ge ... \ge \lambda_n$, where $\overline{\boldsymbol{X}}$ is the mean value of each one of the $n$ original features, eq.\eqref{eq:xmean}, and $m$ is the total number of entries in the data set. The projection of the data onto the principal component space ensures a maximal variance on the direction of the first component. Each subsequent principal component is orthogonal to the previous ones and points in the direction of maximal variance in the residual sub-space \cite{Shlens2014}.
	
	\begin{align}
	\Sigma_{ij} & = \frac{1}{m} \sum_{k=1}^{m} \left( \boldsymbol{X}_i^k - \overline{\boldsymbol{X}_i} \right)\left( \boldsymbol{X}_j^k - \overline{\boldsymbol{X}_j} \right) \label{eq:covariance} \\
	\overline{\boldsymbol{X}} & = \frac{1}{m} \sum_{i=1}^{m} \boldsymbol{X}_i \label{eq:xmean} \\
	\boldsymbol\Sigma \boldsymbol v & = \boldsymbol\lambda \boldsymbol v
	\end{align}
	
	The PCA transformation creates the same number of components in the transformed space, $\boldsymbol{\tilde{X}}$, as features in the original data space $\boldsymbol{X}$. However, components with small eigenvalues belong to a dimension where the variance is so small that it is impossible to separate points in the data. It is a general practice in data reduction to keep only the first $k$ components that explain at least a significant portion of the total variance of the data, $\sum_{i=1}^k \lambda_i /\text{Tr}(\Sigma) > \epsilon$. This allows for a selection of information that will effectively differentiate data points, and for a reduction of the amount of data to process during analysis. Many techniques have been suggested for the selection of the values of $k$ and the cut-off $\epsilon$ \cite{Rea2016}. We use the value of $\epsilon=0.95$.
	
	\subsubsection{Dimension Reduction using Kernel PCA}
	\label{sec:reduckpca}
	PCA has a limitation: the principal components are a linear combination of the original properties of the solar wind. The Kernel PCA (KPCA) is an extension of the PCA that allows to perform non-linear transformations of the original data. The goal in KPCA is to perform the original PCA operations in a high dimensional space.
	
	For a list of $m$ data points composed of $n$ features, it is sometimes difficult (or impossible) to build a linear hyper-plane that dissects regions of different density. However, it is possible to conceive a function, $\boldsymbol{\xi}: \boldsymbol{X}\in\mathbb{R}^n \rightarrow \boldsymbol{\tilde X}\in\mathbb{R}^m$, that will transform all the data into a space where each cluster of points can be linearly separable. The goal is then to avoid explicitly calculating the high-dimensional function $\boldsymbol{\xi}$ by building a Kernel, $\boldsymbol{K}$, which is the inner product of the high-dimensional space:
	
	\begin{align}
	\boldsymbol{K}=k(\boldsymbol{X_i},\boldsymbol{X_j}) = \boldsymbol{\xi}(\boldsymbol{X}_i)^T \boldsymbol{\xi}(\boldsymbol{X}_j)
	\end{align}
	
	In this space the projected points are linearly separable using the same principles of the PCA. In this case the covariance matrix would be expressed as:
	
	\begin{align}
	\Sigma_{ij} & = \frac{1}{m} \sum_{k=1}^{m} \boldsymbol{\xi}(\boldsymbol{X_i})\boldsymbol{\xi}(\boldsymbol{X_i})^T \label{eq:covariancekpca} \\
	\boldsymbol{v} &= \sum_{i=1}^n a_i \boldsymbol{\xi}(x_i) \label{eq:eigphi}
	\end{align}
	
	Popular kernel functions include Gaussian, polynomial and hyperbolic tangent. The transformation is reduced to solving the eigenvalue problem: $\boldsymbol{K} \boldsymbol{a} = \boldsymbol{\lambda}\boldsymbol{a}$, where $\boldsymbol{a}$ are the coefficients of the linear combination of the eigenvectors, eq.\eqref{eq:eigphi}. Although a powerful tool, KPCA requires the creation of an $m\times m$ matrix that can consume large amounts of time and memory resources.
	
	In this work we use KPCA with a polynomial kernel of order eight (8). We also apply the procedure described before to select the total number of retained components: we impose $\epsilon = 0.95$. Cutting off the number of components implies a loss of data. To verify that only minimal information is lost, we perform a transformation of all our data set followed by an inverse transformation. The relative error between the two is normally distributed around zero with less than 1\% of variance.
	
	\subsubsection{Dimension Reduction Using Autoencoders}
	\label{sec:reducae}
	An alternative to data reduction is the use of Autoencoders (AE). These are machine learning techniques that can create non-linear combinations of the original features projected on a latent space with less dimensions \cite{Hinton2006}. This is accomplished by creating a system where an encoding function, $\phi$, maps the original data $\boldsymbol{X}$ to a latent space, $\boldsymbol{\mathcal{F}}\in\mathbb{R}^d$, eq.\eqref{eq:encoder}. A decoder function, $\psi$, then maps the latent space back to the original input space, eq.\eqref{eq:decoder}. The objective of the autoencoder is to minimize the error between the original data and the data produced by the compression-decompression procedure as shown in eq.\eqref{eq:aeminimization}.
	
	\begin{align}
	\phi: & \boldsymbol{X} \rightarrow \boldsymbol{Z}\in\boldsymbol{\mathcal{F}} \label{eq:encoder}\\
	\psi: & \boldsymbol{Z}\in\boldsymbol{\mathcal{F}} \rightarrow \boldsymbol{X} \label{eq:decoder} \\
	\phi,\psi = & \underset{\phi,\psi}{\arg \min} \left\lVert \boldsymbol{X} - (\phi \circ \psi) \boldsymbol{X} \right\rVert^2 \label{eq:aeminimization}
	\end{align}
	
	Autoencoders can be represented as feed-forward neural networks, where fully connected layers lead to a central bottleneck layer with few nodes and then expands to reach again the input layer size. An encoded element, $z \in \boldsymbol{\mathcal{F}}$, can be obtained from a data entry, $x \in \boldsymbol{X}$, following the standard neural network function, eq.\eqref{eq:encodex}, where $\boldsymbol{W}$ is the weights matrix, $c$ is the bias, and $f$ is the non-linear activation function.
	
	\begin{align}
	z & = f \left( \boldsymbol{W}x + c \right) \label{eq:encodex} \\
	\hat{x} & = f' \left( \boldsymbol{W'}z + c' \right) \label{eq:decodez} \\ 
	\mathcal{L}(x, \hat{x}) & =  \left\lVert x - \hat{x} \right\rVert^2 \label{eq:aeloss}
	\end{align}
	
	The decoding procedure, shown in eq.\eqref{eq:decodez}, transforms $z\rightarrow\hat{x}$, where the prime quantities are associated with the decoder. The loss function, $\mathcal{L}(x,\hat{x})$, is the objective to be minimized by the training of the neural network using gradient descent. Once training is completed, the vector $z$ is a projection of the input vector $x$ onto the lower dimensional space $\boldsymbol{\mathcal{F}}$.
	
	Additional enhancements and variations of this simple autoencoder setup exist in the literature, including multiple regularization techniques to minimize over-fitting \cite{7407967}, Variational Autoencoders (VAE) that produce encoded Gaussian distribution functions \cite{Kingma2013}, and Generative Adversarial Networks that automatically generate new data \cite{Goodfellow2014}. In this work we use the most basic form of autoencoders, presented above.
	
\begin{figure}[h!]
	\begin{center}
		\includegraphics[width=18cm]{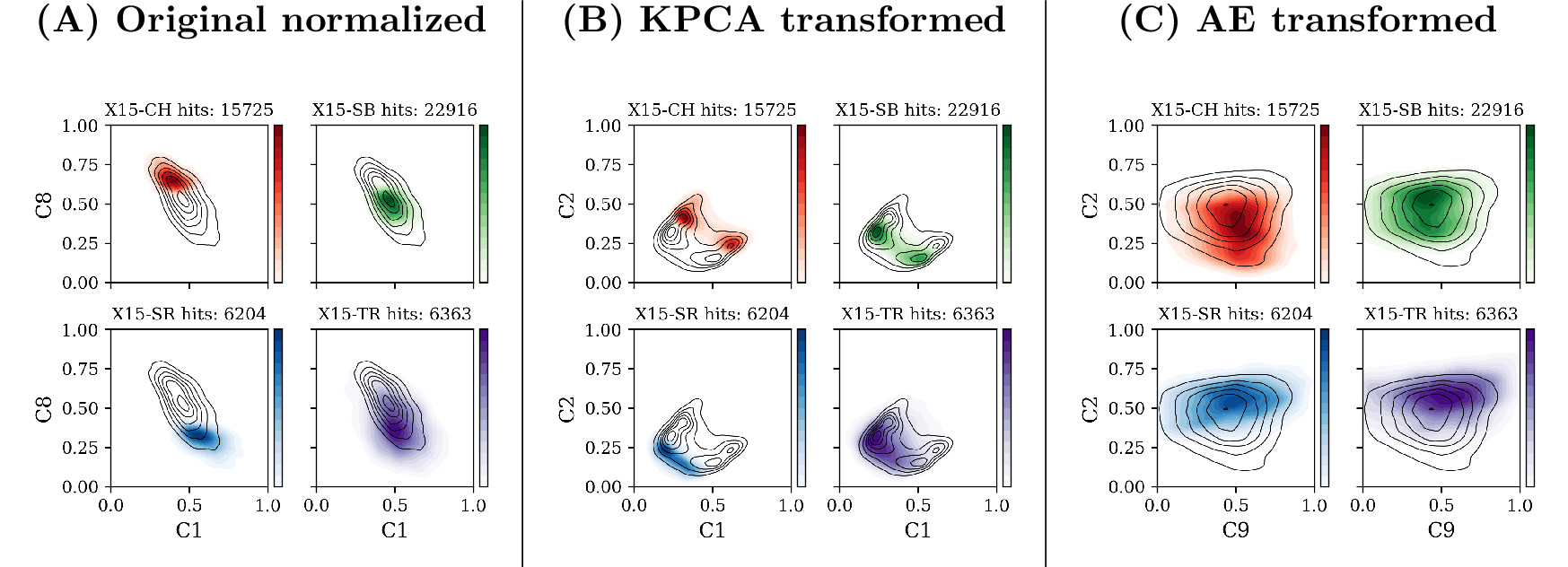}%
	\end{center}
	\caption{Density of points, projected on two arbitrary components, for each of the four X15 classes. The axis titles indicate the corresponding component (C1 for component 1, C8 for component 8, etc.). Colors, normalized between 0 and 1, correspond to solar wind classes: CH (red), SB (blue), SR (green), and TR (purple). The three columns correspond to three possible data transformations: (A) Original normalized data, (B) data transformed with the Kernel Principal Component Analysis, and (C) data encoded with our Autoencoder. Black lines are isocontours of data point density.}\label{fig:dimreduc}
\end{figure}

	In the present work we will be showing different representations of the solar wind data, transformed with different techniques and projected on flat planes. Fig.\ref{fig:dimreduc} presents our data set in three different projections: (A) the original feature space, normalized between zero and one, (B) the transformed data set using the KPCA method, and (C) the AE transformed data. In each panel four histograms present the distribution of the X15 classes, on two arbitrary components identified by the axis title.
	
	\subsubsection{Clustering Techniques}
	\label{sec:clustering}
\begin{figure}[h!]
	\begin{center}
		\includegraphics[width=18cm]{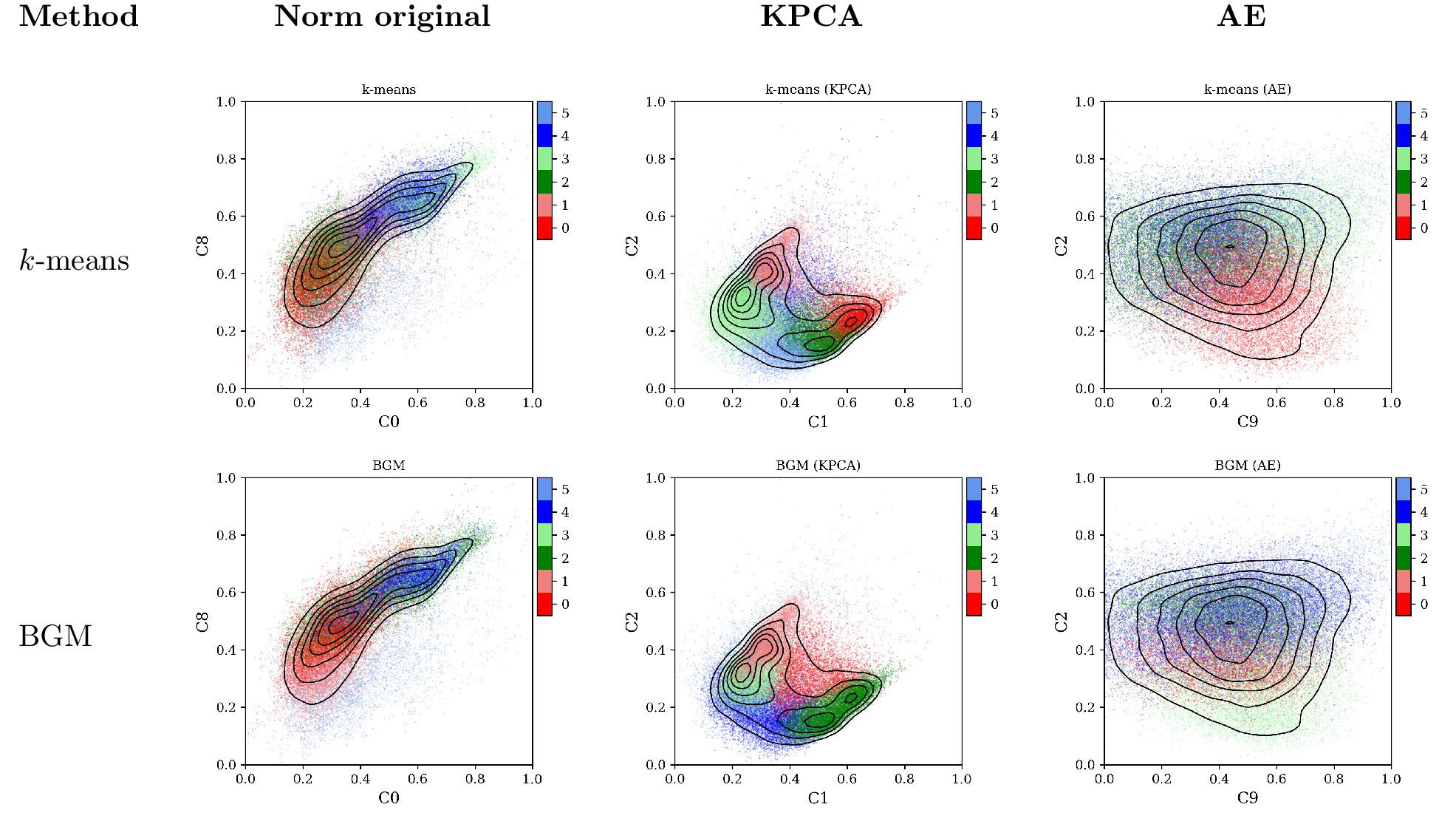}%
	\end{center}
	\caption{Scatter plot of all data points projected on two of the components of the three transformed spaces: the original normalized space (left column), the KPCA space (central column), and the AE space (right column). In the left column the components C0 vs C8 correspond to the proton temperature, $T_p$, vs the proton specific entropy, $S_p$. Colors correspond to the classes obtained by two unsupervised clustering methods: $k$-means (first row) and BGM (second row). Black lines are isocontours of data point density.}\label{fig:clustering}
\end{figure}

	The goal of unsupervised machine learning is to group data points in a limited number of clusters in the N-dimensional space $\Omega\in\Bbb R^n$, where $n$ is the number of features (components or properties) in the data set. Multiple techniques can be used to perform multi-dimensional clustering. We present in Fig. \ref{fig:clustering} the application of two basic clustering techniques to classify our data set. Following the same order as before, the first column in the figure contains all data points projected in the original normalized feature space; column two contains scatter plots of the points after KPCA transformation; column three contains the same points encoded in the AE latent space. Each row corresponds to a different clustering method. The colors in the top row were obtained using the $k$-means method \cite{1056489}, while the colors in bottom panels were obtained using the Bayesian Gaussian Mixture (BGM) \cite{bishop2006machine}.
	
	The $k$-means technique has already been used in multiple publications for the determination of solar wind states \cite{Heidrich-Meisner2018,Roberts2020}. The BGM technique has also been recently used by \cite{Bloch2020} to classify solar wind observations by the ACE and Ulysses missions. Mixture models similar to the BGM have also been recently used to classify space plasma regions in magnetic reconnection zones \cite{Dupuis2020}. None of these previous publications used data transformation to solve the classification problem in a more suitable latent space.
	
	The colors used in Fig.\ref{fig:clustering} are assigned randomly by each clustering technique. The most glaring issue with them is that different methods can lead to different clusters of points. The BGM and the $k$-means do not agree on their classification in the PCA and the AE space. More importantly, for each technique, slight modifications of the clustering parameters, e.g. using a different seed for the random number generator, can lead to very different results. We address this last issue using an algorithm that launches the $k$-means (the BGM) algorithm 100 (30) times until the method converges to a global minimum. The final results are implementation dependent.
	
	In the present data set, the cloud of points is convex and well distributed in all components. This raises one additional issue, observed more clearly in the second column of Fig.\ref{fig:clustering}: when classical clustering methods are applied to relatively homogeneously dense data, it divides the feature space in Vorono\"i regions with linear hyper-plane boundaries. This is an issue with all clustering techniques based on discrimination of groups using their relative distances (to a centroid or to the mean of the distribution). To avoid this problem density-based techniques, such as DBSCAN \cite{ester1996density}, and agglomeration clustering methods, use a different approach. However, we can not apply them here because in such homogeneous cloud of points these techniques lead to a trivial solution where all data points are assigned to a single class. An alternative projection was used by \cite{Bloch2020}, who performed a Uniform Manifold Approximation and Projection (UMAP). We performed the same projection unsuccessfully: the Ulysses data used in that publication contains a very dense and large number of CH observations. ACE lacks such a rich variety of CH data, so applying a UMAP leads to a single class.
	
	There is no guarantee that a single classification method, with a particular set of parameters will converge to a physically meaningful classification of the data if the points in the data do not have some level of separability, or have multiple zones of high density. This is also true for other classification methods based on \textit{supervised learning}. The same issues will be observed when the training data include target classes derived from dense data clouds using simple hyper-plane boundaries, as done for the Z09 and X15 classes. An example of such application was published by \cite{Camporeale2017b,Li2020}. The authors used the X15 classification to train supervised classifiers. No new information is gained with such methods, as the empirical boundaries are already mathematically known. A more compeling task would be to compare all classification methods against a \textit{ground truth}, i.e. against a catalog of footpoint locations on the solar surface. But such catalog, to our knowledge, does not exist.
	
	\subsubsection{Self-Organizing Maps}
	\label{sec:som}
	
	\paragraph{Classical SOM}
	\label{sec:classicalsom}
	
	Following the definitions and notations by \cite{Villmann2006}, a class can be defined as $C_i\overset{\text{def}}{=} \{x\in\Omega | \Phi(x)=\boldsymbol{w}_i\}$, where $\Phi$ is a function from $\Omega$ to a finite subset of $k$ points $\{\boldsymbol{w}_i\in\Bbb R^N\}_{i=1..k}$. A cluster $C_i$ is then a partition of $\Omega$, and $\{\boldsymbol{w}_i\}$ are the code words (also known as nodes, weights or centroids) associated. The mapping from the data space to the code word set, $\Phi: \Omega\rightarrow\mathcal{W}$, is obtained by finding the closest neighbor between the points $\boldsymbol{x}$ and the code words $\boldsymbol{w}$, eq.\eqref{eq:winner}. The code word $\boldsymbol{w_s}$, the closest node to the input $\boldsymbol{x_s}$, is called the \textit{winning element}. The class $C_i$ corresponds to a Vorono\"i region of $\Omega$ with center in $\boldsymbol{w}_i$.
	
	\begin{equation}
	\Phi: x \rightarrow  \underset{i\in\mathcal{N}}{\arg\min}\left( \left\lVert x - \boldsymbol{w}_i \right\rVert \right) \label{eq:winner}
	\end{equation}
	
	A Self-Organizing Map (SOM) is also composed of structured nodes arranged in a lattice, each one assigned to a fixed position $\boldsymbol{p}_i$ in $\Bbb R^q$, where $q$ is the dimension of the lattice (generally $q=2$). The map nodes are characterized by their associated code words. The SOM learns by adjusting the code words $\boldsymbol{w}_i$ as input data $x$ is presented.
	
	The SOM is the ensemble of code words and nodes $\{ \boldsymbol{w}_i, \boldsymbol{p}_i\} \in (\Omega \times \Bbb R^q)$. For a particular entry $\boldsymbol{x_s}$, the code word $s \in \mathcal{N}$ is associated to the winning node $\boldsymbol{p_s}$ if the closest word to $\boldsymbol{x_s}$ is $\boldsymbol{w_s}$. At every iteration of the method, all code words of the SOM are shifted towards $x$ following the rule:
	
	\begin{equation}
	\Delta \boldsymbol{w}_i = \epsilon(t)h_\sigma(t,i,s)(x-\boldsymbol{w}_i) \label{eq:learnsom}
	\end{equation}
	
	with $h_\sigma(t,i,j)$ defined as the lattice neighbor function:
	
	\begin{equation}
	h_\sigma(t,i,j) = e^{-\frac{\left\lVert \boldsymbol{p}_i - \boldsymbol{p}_j \right\rVert^2}{2\sigma(t)^2}} \label{eq:neigsom}
	\end{equation}
	
	where $\epsilon(t)$ is the time dependent learning rate, eq.\eqref{eq:epsilon}, and $\sigma(t)$ is the time dependent lattice neighbor width, eq.\eqref{eq:sigma}. The training of the SOM is an iterative process where each data point in the data set is presented to the algorithm multiple times $t={0, 1,..,t_f}$. In these equations the subscript $0$ refers to initial values at $t=0$ and the subscript $f$ to values at $t=t_f$.
	
	\begin{align}
	\epsilon(t) & = \epsilon_0 \left(\frac{\epsilon_f}{\epsilon_0}\right)^{t/t_f} \label{eq:epsilon} \\
	\sigma(t) & = \sigma_0 \left(\frac{\sigma_f}{\sigma_0}\right)^{t/t_f} \label{eq:sigma}
	\end{align}
	
	This procedure places the code words in the data space $\Omega$ in such a way that neighboring nodes in the lattice are also neighbors in the data space. The lattice can be presented as a $q$-dimensional image, called map, where nodes sharing similar properties are organized in close proximity.
	
	The main metric for the evaluation of the SOM performance is called the quantization error:
	
	\begin{equation}
	Q_E = \frac{1}{m} \sum_{i=1}^m \left\lVert x_i - w_{x_i} \right\rVert \label{eq:QE}
	\end{equation}
	
	where $m$, is the total number of entries in the data set. It has been shown that the SOM tends to converge in the mean-square (m.s.) sense to the probabilistic density center of the multi-dimensional input subset \cite{Yin1995}. This means that, if the SOM hyper-parameters are chosen correctly, the code words of the SOM will have a tendency to move towards high density regions of subsets of the input data, and will be located close to the mean of the subset points.
	
	Once the training of the SOM is finished, the code words $\boldsymbol{w}_i$ can be grouped together using any clustering technique, e.g. $k$-means. The nodes of the SOM with close properties will be made part of the same class. The classes created are an ensemble of Vorono\"i subspaces, allowing a complex non-linear partitioning of the data space $\Omega$.
	
	The final number of clusters is an input of the algorithm, but can also be calculated autonomously. The Within Cluster Sum of Squares (WCSS) can be used as a metric of the compactness of the clustered nodes. As its name implies the WCSS is the sum of the squared distances from each node to their cluster point. If only one class is selected, the large spread of the nodes would produce a high WCSS. The lowest possible value of the WCSS is obtained for a very high number of classes, when the number of classes is equal to the number of nodes. But such extreme solution is also unpractical. The optimal number of clusters can be obtained using the Kneedle class number determination \cite{5961514}. We use this automatic technique to let the machine select the optimal number of solar wind classes.
	
	\paragraph{Dynamic SOM}
	The time dependence of the SOM training allows the code words $\boldsymbol{w}_i$ to reach steady coordinates by slowing down their movement over the iterations. Due to the minimization of the distance in eq.\eqref{eq:winner} code words tend to agglomerate around high density zones of the feature space. The Dynamic Self-Organizing Map (DSOM), introduced by \cite{Rougier2011}, eliminates the time dependence and allows to cover larger zones of the space outside of the high density regions.
	
	The DSOM is a variation of the SOM where the learning function \eqref{eq:learnsom} and the neighbor function \eqref{eq:neigsom} are replaced by eqs. \eqref{eq:learndsom} and \eqref{eq:neigdsom} respectively:
	
	\begin{align}
	\Delta \boldsymbol{w}_i & = \epsilon \left\lVert x - \boldsymbol{w}_i \right\rVert_\Omega h_\eta(i,s,x)(x-\boldsymbol{w}_i) \label{eq:learndsom} \\
	h_\eta(i,s,x) & = e^{-\frac{1}{\eta^2}\frac{\left\lVert \boldsymbol{p}_i - \boldsymbol{p}_j \right\rVert^2}{\left\lVert x - \boldsymbol{w}_s \right\rVert_\Omega^2}} \label{eq:neigdsom} 
	\end{align}
	
	where $\epsilon$ is a constant learning rate, $h_\eta(i,s,x)$ is defined as the new lattice neighbor function, and $\eta$ is the \textit{elasticity} parameter. In their work \cite{Rougier2011} show that DSOM can be used to draw a larger sample of the feature space $\Omega$, reducing the agglomeration of code words around high density zones. The main parameters of the DSOM, $\eta$ and $\epsilon$, control the convergence of the method. A large $\epsilon$ moves the code words, $\boldsymbol{w}$, very fast with each new iteration; a very low value moves the points slowly in the space. A high elasticity, $\eta$, keeps all the nodes extremely close to each other, while a low value does not induce movement on far away code words. The best compromise is to use a very low value of the learning rate coupled with a mid-range elasticity, and a large number of training epochs. This can ensure a relative good convergence to a steady set of code words.
	
	One special advantage of the DSOM is that it can be trained \textit{online}, i.e., it is not necessary to re-train all the model when new data arrives: it adapts automatically to new information.
	
	\paragraph{Visualization of SOM and DSOM}
	Most clustering techniques do not guarantee to converge to a steady immutable solution. Differences in the training parameters or slight changes in the data can have an important impact on the final classification. Clustering tools can be used for statistical analysis, comparisons, data visualization and training of supervised methods. But it will be practically impossible to claim the existence of a general objective set of states discovered only by the use of these basic clustering techniques.
	
	However, SOMs and DSOMs provide an important tool for the study of the solar wind: the maps are composed of nodes that share similar properties with its immediate neighbors. This allows for visual identification of patterns and targeted statistical analysis.
	
	We used the python package \href{https://github.com/JustGlowing/minisom}{MiniSom} \cite{vettigli2013minisom} as the starting point of our developments. Multiple methods of the MiniSom have been overloaded to implement the DSOM, and to use a lattice of hexagonal nodes. All auxiliary procedures used to calculate inter-nodal distances, node clustering, data-to-node mapping, and class boundary detection have been implemented by us. All visualization routines are original and have been developed using the python library Matplotlib \cite{hunter2007matplotlib}.
	
\begin{figure}[h!]
	\begin{center}
		\includegraphics[width=12cm]{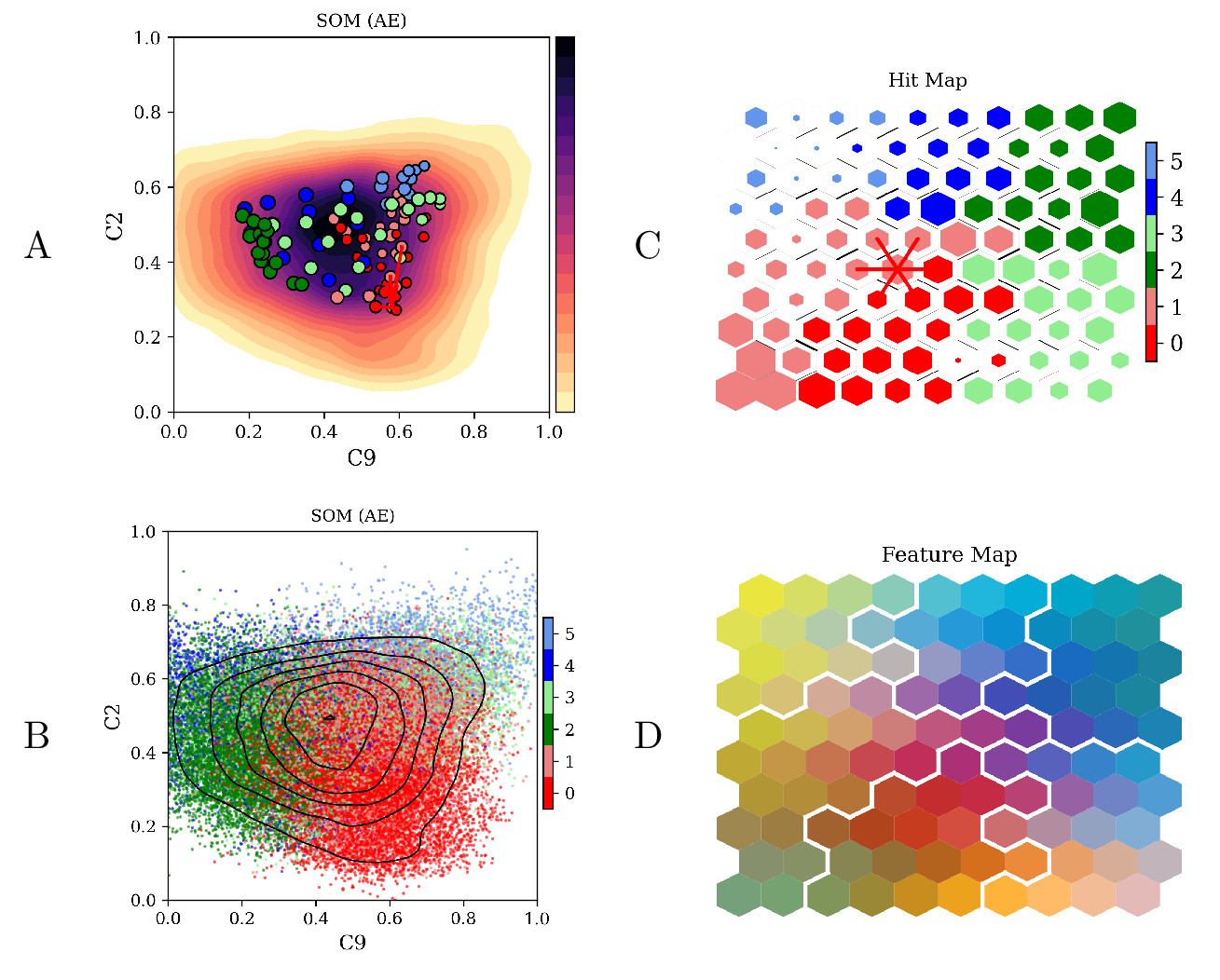}%
	\end{center}
	\caption{Visualization of the Self-Organizing Maps. Panel (A): histogram with the normalized density of data points superposed by the code words of the DSOM, projected on two components of the latent AE space. A single node is connected to its closest neighbors by red lines. Panel (B): scatter plot of all data points, colored by the DSOM class. Panel (C) Hit map: the size of the hexagon corresponds to the number of data points associated to the map node, and the color is the corresponding DSOM class. Black lines between nodes represent their relative distance. Red lines connect the nodes similarly highlighted in panel (A). Panel (D): Map of the nodes colored by three of their components, combined as a single RGB color. White lines mark the boundaries between DSOM classes.}\label{fig:maps}
\end{figure}

	Fig.\ref{fig:maps} shows the basic types of plots that can be generated using the SOM/DSOM techniques. We present in this figure the outcome of our model, combining a non linear AE transformation of the ACE data set with the unsupervised classification of the encoded data using the DSOM method. Panel (A) shows a histogram of two components of the feature space $\Omega$, with dots marking the position of the code words $\boldsymbol{w}_i$. The colors of the dots represent their DSOM classification. The red lines connect a single code word $\boldsymbol{w}_s$ with its six closest neighbors. Panel (B) shows the same information as in the previous panel, but using a scatter plot colored by the DSOM classification. This image shows the domain of influence of each one of the DSOM classes.
	
	Panel (C) shows the \textit{hit map} of the DSOM. It contains the lattice nodes $\boldsymbol{p}_i$ associated to the code words $\boldsymbol{w}_i$. They are depicted as hexagons with sizes representing the number of data points connected to each node and colored by their DSOM class. The thickness of the lines between lattice nodes represent the relative distance to its neighbors in the feature space $\Omega$. Red lines connect the node $\boldsymbol{p}_s$, associated to the code word $\boldsymbol{w}_s$ in panel (A), to its closest neighbors.
	
	Panel (D) of Fig.\ref{fig:maps} displays three components of the code words $\boldsymbol{w}_i$ associated to each one of the $\boldsymbol{p}_i$ nodes. The node components have been mapped to the basic colors Red, Green and Blue (RGB) and combined together to produce the composite color shown in the figure.
	
	These four representations are only a few examples of the variety of data that can be represented using SOMs. The most important aspect of the SOMs is that data is represented in simple 2D lattices where the nodes share properties with their neighbors. Here we also decided to use hexagonal nodes, connecting 6 equidistant nodes, but other types of representations are also valid, e.g. square or triangular nodes.
	
	\subsubsection{The Full Architecture}
	\label{sec:fullarchi}
\begin{figure}[h!]
	\begin{center}
		\includegraphics[width=16cm]{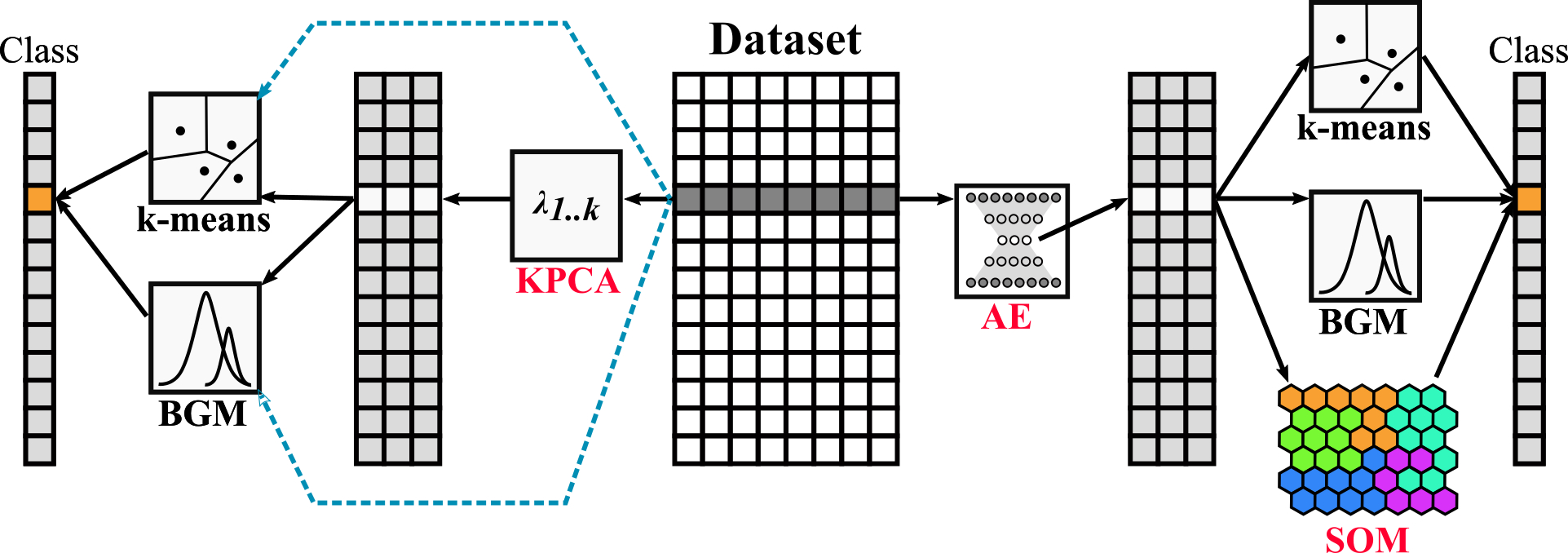}%
	\end{center}
	\caption{General overview of the pipelines tested in this work. Starting from the center, the ACE data set is processed and normalized. Blue dashed lines show the work done in previous publications by different authors. Black lines show how data in this work is first transformed and then classified using multiple methods. The original techniques presented in this paper are highlighted in red.}\label{fig:architecture}
\end{figure}

	The previous sections introduced all the individual pieces that we use for the present work. Here, we give a global view of the full model. Fig.\ref{fig:architecture} shows how all the components are interconnected. At the center of the image is the processed and normalized original ACE data set. The blue dashed lines show the unsupervised techniques already presented by \cite{Bloch2020,Roberts2020,Heidrich-Meisner2018}. The KPCA step is added to the data pipelines used in the literature in order to project the data into a hyper-space where the class boundaries are better defined.
	
	On the right side of the same figure we present our main approach: we perform first a data encoding using an AE, then we perform unsupervised classification of the solar wind with the $k$-means, BGM and DSOM methods. After training, the code words of the DSOM are clustered to group together nodes that share similar properties. This second level classification is done using the $k$-means++ algorithm with 100 re-initializations (it is in general recommended to use between 50 and 500 initializations, searching for a global optimum, as different random runs can lead only to a local minima). We use the Kneedle method to automatically select the number of classes that the DSOM will produce \cite{5961514}. The BGM and the $k$-means clustering techniques are included for comparison.
	
	All the software was implemented in Python using as main libraries PyTroch \cite{paszke2019pytorch}, Scikit-learn \cite{pedregosa2011scikit}, Matplotlib \cite{hunter2007matplotlib}, MiniSom \cite{vettigli2013minisom}, Pandas \cite{mckinney2010} and NumPy \cite{oliphant2006guide}.
	
	\paragraph{Feature selection}
	\label{sec:featureselection}
	Table \ref{tab:features} lists all the features used in our model. A detailed description of each feature can be found in the \href{http://www.srl.caltech.edu/cgi-bin/dib/rundibviewmultil2/ACE/ASC/DATA/level2/multi}{ACE Level 2 documentation}. To spread the data over a larger range of values in each component, we have used the logarithm of all the quantities, except of those marked with an asterisk in the table.
	
	Features 11 to 15 contain an additional suffix, corresponding to a statistical operation performed on the corresponding feature. In our model we only include \textit{range} operations, but we have provided our software with the ability to calculate also the mean, the standard deviation and the auto-correlation of quantities over a window of time of 6 hours. This window allows to capture temporal (spatial) fluctuations in some of the solar wind parameters.
	
	On the lower part of Table \ref{tab:features} we present the range of dates used for the model. The same table also contains the hyper-parameters selected to run the two models. The number of neurons per layer in the encoding half of the neural network is listed in the table.
	
	\paragraph{Autoencoder architecture}
	We use a basic, fully connected feed-forward neural network for the encoding-decoding process. The neural network is symmetric in size but the weights of the encoder, $\boldsymbol{W}$, and the decoder, $\boldsymbol{W'}$, are not synchronized (see eqs.\eqref{eq:encodex}, \eqref{eq:decodez}). Each layer is composed of a linear regressor, followed by a GELU activation function. The output layer of the network contains a linear regressor followed by a sigmoid activation function. The AE has been coded in python using the PyTorch framework \cite{paszke2019pytorch}.
	
	The final architecture of the AE and its hyper-parameters have been optimized automatically using the \textit{Optuna} library \cite{Akiba2019}. We instructed this Hyper-Parameter Optimization (HPO) to select the optimal values for the following parameters, given the corresponding constraints:
	
	\begin{itemize}
		\item \textbf{Number of layers}: an integer between 2 and 6.
		\item \textbf{Number of neurons per layer}: it must be larger than 3 and smaller than the number of neurons in the previous encoder layer.
		\item \textbf{The neural network optimizer}: selected among Adam, Stochastic Gradient Descent, and RMSprop.
		\item \textbf{The learning rate}: a float value between 10$^{-5}$ and 10$^{-1}$.
	\end{itemize}
	
	The automatic HPO is based on a technique called Tree-structured Parzen Estimator (TPE) \cite{pmlr-v28-bergstra13}, which uses Bayesian Optimization to minimize a target function, $\mathcal{H}$, provided by the user. We use the test loss of the AE as target function to be minimized.
	
	The HPO performs a total of one thousand (1000) different trials. However, to accelerate the optimization process, we built a smaller complementary data set. To avoid over-fitting on a sub-set of the original data we used the $k$-means algorithm to produce a representative sample of  $m'=4\sqrt{m}$ data points. This allows to explore a much broader set of hyperparameters in a short period of time. This artificial data set is then discarded and the AE is trained on the real data set.
	
	The HPO selected the Adam optimizer \cite{Kingma2014} for the gradient descent with a learning rate of 0.042. The total number of layers selected is 2, and the number of nodes in the bottleneck is 10. The loss function is the Mean Squared Error (MSE). We train the network for 500 epochs, after which no additional improvement in the loss function is observed. The full data set was randomly divided 50\%/50\% between training and testing sets. We track the evolution of both data sets during training. We did not observe any variance or bias error.
	
\begin{figure}[h!]
	\begin{center}
		\includegraphics[width=16cm]{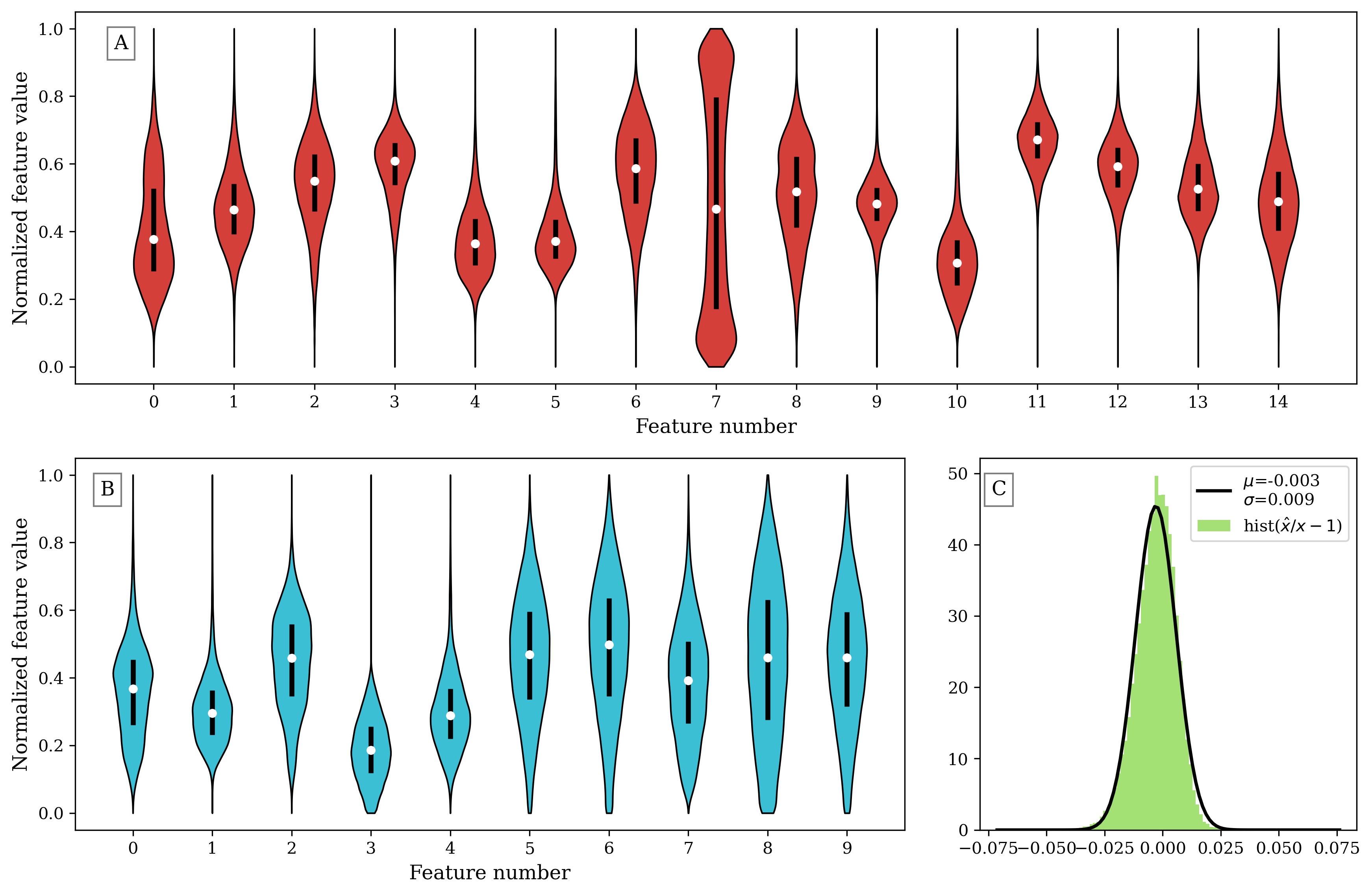}%
	\end{center}
	\caption{Violin plots showing the data distribution in (A) the original normalized data set, and (B) the AE transformed data set. Panel (C) shows a histogram of the relative error produced by the lossy compression-decompression procedure in the AE. The error is close to zero, with a variance of less than 1\%.}\label{fig:datarange}
\end{figure}

	The final architecture is trained using the full data set for 500 epochs. Fig.\ref{fig:datarange} shows the distribution of data in the original feature space, panel (A), and in the AE latent space, panel (B). The data in the original space contains extreme data points far from the mean value, and most features present a normal distribution. The combination of these two properties makes it difficult for any unsupervised clustering technique to separate points and accurately categorize different kinds of solar wind.
	
	Panel (C) shows the error in the encoding-decoding procedure of the AE. It shows a histogram of the relative error, $\mathbf{E_r} = \mathbf{\hat{X}}/\mathbf{X} - 1$, observed between the input data, $\mathbf{X}$, and the decoded values, $\mathbf{\hat{X}}$. A normal distribution function has been fitted to the values of the histogram. It shows that the relative error is centered near zero and its variance is around 1\%. 
	
	\paragraph{Selection of parameters for the DSOM}
	
	In this manuscript we have introduced the use of the DSOMs for the classification of solar wind data. This technique requires the selection of four main Hyper-Parameters (HPs): the size of the lattice, $(L_x\times L_y)$, the constant learning rate, $\epsilon$, and the elasticity, $\eta$. These last two parameters where chosen manually, while the lattice size was automatically selected by Hyper-Parameter Optimization (HPO) using \textit{Optuna} \cite{Akiba2019}.
	
	For the selection of the number of nodes in the lattice we propose the use of the objective function, $\mathcal{H}$, described in eq.\eqref{eq:hpo}:
	
	\begin{equation}
	\mathcal{H}\left(\sigma, \eta, L_x, L_y\right) = \frac{Q_E(\sigma, \eta, L_x, L_y)}{Q_0} + \alpha \frac{L_x}{m_{\max}} + \beta \frac{L_y}{n_{\max}} + \gamma \frac{L_x L_y}{\max\left( {m_{\max},n_{\max}} \right)} \label{eq:hpo}
	\end{equation}
	
	where $Q_E$ is the quantization error at the end of the training, $Q_0$ is a reference quantization error before training, $L_x$ and $L_y$ are the number of lattice nodes in each dimension, and $m_{\max}$ and $n_{\max}$ are the given maximum number of possible nodes. The weight factors $\alpha$, $\beta$ and $\gamma$ are used to impose restictions on each term. We have fixed their value to $\alpha=\beta=0.08$ and $\gamma=0.6$. When a large number of nodes is available smaller values of $Q_E$ are automatically obtained because the mean distance from the data set entries to the code words is reduced. The second and third terms on the RHS of $\mathcal{H}$ leads the optimizer to reduce the number of nodes in the SOM. The squaring term $\gamma L_x L_y$ forces the map to be as squared as possible.
	
	After a total of 500 trial runs of the model using different HPs, the optimizer selected the parameters presented in the lower section of table \ref{tab:features}. The optimization was accelerated using the same technique as in the optimization of the AE: we generated a reduced number of points using the $k$-means algorithm, with a total number of entries equal to one twentieth the size of the full data set, $m'=\frac{1}{20} m$.
	
	The two remaining parameters of the DSOM, the elasticity $\eta=3.0$ and the learning rate $\epsilon=0.005$, have been manually selected. These two values control the speed at which the code words move towards the data entries, and the attraction between neighboring code words. It has been shown by \cite{Rougier2011} that high values of the elasticity, $\eta$, lead to tightly packed code words, while low values lead to loose connections. The elasticity takes in general values between 1 and 10. On the other hand, the learning rate indicates to the code words how fast they should move towards new incoming data. Very small learning rates could lead to very slow convergence to a solution, while very large values might produce code words that jump from value to value without converging to a global solution. The value of the learning rate can be set somewhere between 0.001 and 0.9.
	
\begin{figure}[h!]
	\begin{center}
		\includegraphics[width=16cm]{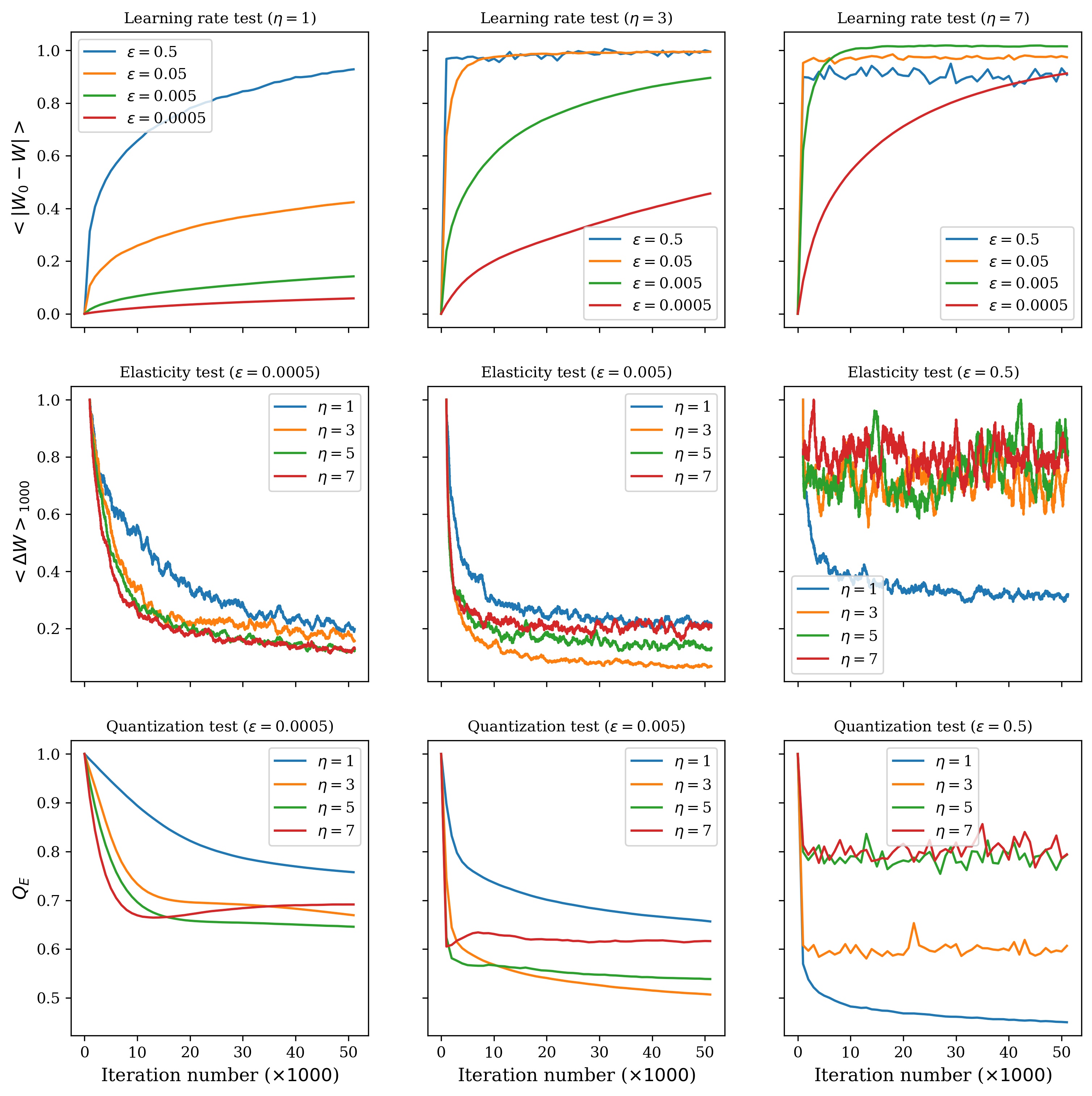}%
	\end{center}
	\caption{Effects of the elasticity, $\eta$, and the learning rate, $\epsilon$, on the training of the DSOM. Top row: mean value of the difference between the position of the code words at each iteration, $W$, and their original position, $\left<\lvert W_0 - W \rvert \right>$. Middle row: moving average (1000 iterations) of the mean distance traveled by the code words in one iteration, $\left<\Delta W \right>_{1000}$. Bottom row: quantization error per iteration, $Q_E$.}\label{fig:somstudy}
\end{figure}

	Fig.\ref{fig:somstudy} shows how the elasticity and the learning rate can affect the convergence of the DSOM. In this figure we evaluate the effect of using different values of $\eta$ and $\epsilon$. Three different graphs are used to understand the evolution of the training and its convergence to a stable solution. The first row shows how the code words move away from their original position during the training: as the iterations advance the code words move until they find a stable location. It is clear that lower values of $\epsilon$ and $\eta$, as presented in the left panel of the first row, lead to very long convergence times. At the other extreme, very high values of the two parameters produce strong movements with a compact group of code words, leading to a non-converging solution.
	
	In the second row of the same figure we show the distance traveled by the code words at each iteration of the training. In the best case scenario this distance is large at the beginning of the training and converges towards zero as the iterations pass. The third panel of this row shows how large values of $\eta$ and $\epsilon$ produce solutions of the DSOM that do not converge.
	
	The third row of Fig.\ref{fig:somstudy} shows the evolution of the quantization error, eq.\eqref{eq:QE}. This value explains the compactness of the data points around the code words. Scattered points will show large $Q_E$, while dense clouds of points gathered around the code words will show low $Q_E$ values. Once again in this last row we see that there is a compromise between a slow convergence with small values of $\epsilon$, and large values of the two parameters that can lead to unstable solutions.
	
	This figure also shows that, even if the DSOM is a dynamic technique that does not use a decay of the learning rate with time, it is a method that converges to a steady solution, if the parameters are properly selected.
	
	\paragraph{Budget}
	Machine learning models require fine tuning of different parameters, from the selection and testing of multiple methods, to the parameterization of the final architecture. \cite{Dodge2019} suggests that every publication in machine learning should include a section on the budget used for the development and training of the method. The budget is the amount of resources used in the data processing, the selection of the model hyper-parameters (HP), and its training.
	
	The most time-consuming task in the present work has been the data preparation, the model setup and debugging and the writing of the SOM visualization routines. All the techniques described in the previous sections have been coded in python and are freely accessible in the repositories listed in section \ref{sec:repos}. We estimate the effort to bring this work from scratch to a total of 2 persons month. Of these, one person week was dedicated to the manual testing an selection of different model HPs (autoencoder architecture, feature selection, learning rates, initialization methods, number of epochs for training, selection of data compression method, size of the time windows, etc.).
	
	All classic clustering techniques presented in section \ref{sec:clustering} require only a few lines of code and can be trained in minutes on a mid-range workstation (e.g. Dell Precission T5600, featuring two Intel(R) Xeon(R) CPU E5-2643 0 @ 3.30GHz with four cores and eight threads each). The most time consuming tasks of our models are the training of the autoencoder (5\% of the total run time), the multiple passages of the clustering algorithms (15\% of the run time), and the optimization of the hyper-parameters (80\% of the run time). The training of the DSOM is performed in less than a minute.
	
	For reference, the total run-time of our model is 30 minutes. The python scrips used do not contain any particular acceleration (e.g. using GPUs) or optimizations (e.g. using Numba), so there is large room for improvement of the computational efficiency.
	
	\section{Results and comparisons}
	\label{sec:results}
	\subsection{Interpretation of the DSOM plots}
	
\begin{figure}[h!]\centering
	\includegraphics[width=18cm]{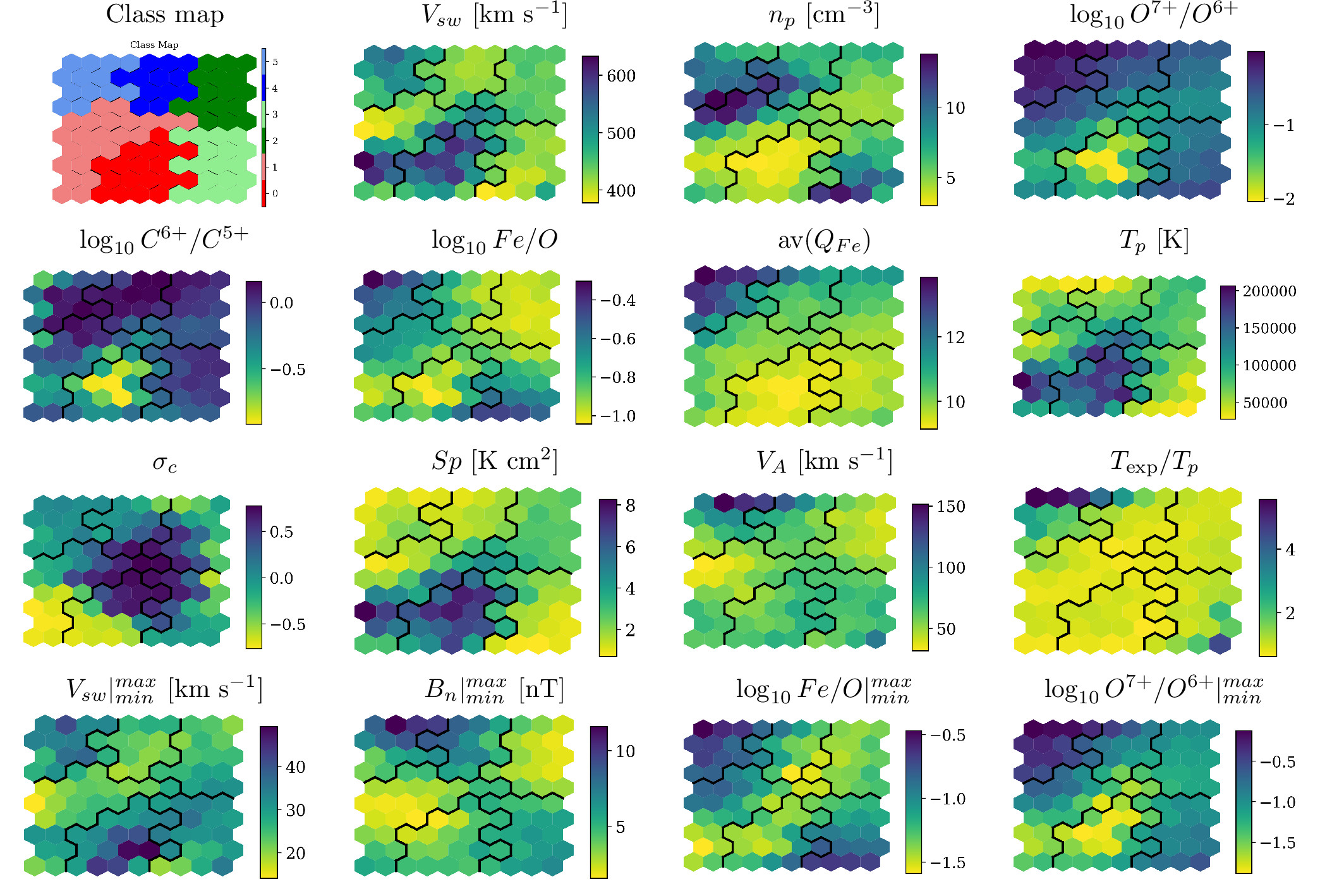}
	\caption{Map colored by the DSOM classes (top left panel), and composition of the solar wind associated to each one of the map nodes. The black line marks the boundaries between DSOM classes.}\label{fig:compmap}
\end{figure}

	When the DSOM method converges to a solution, each one of the code words is a representative of their N-dimensional neighborhood. We perform then a k-means clustering of the code words and apply the Kneedle method \cite{5961514}, presented in section \ref{sec:classicalsom}, to select the final number of classes. Here, the automatic procedure selects a total of 6 classes, numbered from 0 to 5. The \textit{Class Map} on the first panel of Fig.\ref{fig:compmap} shows that all nodes are organized in continuous groups.
	
	The weights of the code words can be decoded and scaled to obtain the corresponding physical properties of the associated solar wind. These physical quantities are plotted in Fig.\ref{fig:compmap} for each one of the solar wind features. 
	
	Black continuous lines in the maps mark the boundary between different DSOM classes. All of the maps show uninterrupted smooth transitions between low and high values, without sudden jumps or incoherent color changes. Inside DSOM classes solar wind properties can present variations. This is an expected consequence of projecting 15 dimensions in a 2D lattice. 
	
	The most obvious class to identify is the DSOM class 0, with clear indications of coronal hole origin. It is characterised by very low values of the O7+/ O6+ and C6+/ O5+  ratios, associated with plasma originating from open magnetic field lines~\cite{Zhao2009, Stakhiv2016} , high wind speed, low proton density, high absolute values of $\sigma_c$ (a sign of Alfvenicity), high proton entropy, high proton temperature and moderately high values of Alfven speed (associate by~\cite{Xu2015b} with coronal holes).
	
	The proton density has a very broad range of values for class 1. A close examination of the map of cross-helicity, $\sigma_c$, shows that this class also contains Alfv\'enic solar wind with both polarities. Class 1 also showcases high proton temperatures, high solar wind speeds, but average oxygen and carbon ionization ratio, and average iron charge. All these observations point towards solar wind originated at the boundary of coronal holes \cite{Zhao2017}.
	
	Class 5 can be associated to transient event, such as ICME and ejecta.  It presents the very high O7+/ O6+  ratio values that \cite{Zhao2009, Stakhiv2016, Xu2015b} associate to CME plasma, and the low proton temperature values usually found in ICMEs. It is also characterised by the high (but, quite surprisingly, lower than for class 0) solar wind velocity, $\sigma_c \sim 0$ \cite{Roberts2020}, the high values of Alfv\'en speed which are usually associated to explosive transient activity~\cite{Xu2015b}.
	
	Class 4 has similar properties as class 5 and can be mainly composed of transient events, but it also contains more Alfv\'enic plasma, and very high carbon charge state ratios, $C^{6+}/C^{5+}$. Fluctuations in this class are slightly less significant than the ones observed in class 5, except for jumps in the normal magnetic field, range $B_n$. These can point towards a class that contains magnetic clouds or Sector Boundary Crossing (SBC) events. Classes 4 and 5, identified as transients, remain rare, as clearly shown in panel (C) of figure \ref{fig:maps}.
	
	At this point is important to remember that a different set of initial conditions or a different number of map nodes could lead to a slightly different repartition of the data, or to a different number of classes. However, points with similar properties will always remain topologically close and the interpretation of a different set of DSOM classes will lead to similar results. This is not necessarily the case with other unsupervised methods, like k-means, as the topological organization of the data is not maintained, so different runs can produce different results for which previous interpretations can not be re-cycled.
	
	Class 2 and 3 are composed of slow, dense solar wind, the kind of wind that~ \cite{Zhao2017} associates to the \textit{Quiet Sun} and that~\cite{Xu2015b} associates to either Streamer Belt (SB) or Sector Reversal (SR) region plasma. As expected for the slow wind, the cross helicity is low, the proton temperature intermediate between the low values associated to ICMEs and the higher values observed in the fast wind,  the proton entropy and the Alfv\'{e}n speed are low~\cite{Xu2015b}. The high $O^{7+}/O^{6+}$  and $C^{6+}/C^{5+}$ ratios (lower only to the values associates to class mappable to transient events, class 4 and 5), point to plasma originating in closed field lines~\cite{Zhao2009, Stakhiv2016}. Of the two classes, class 2 is characterized by lower wind speed, higher density, lower proton temperature, lower entropy.

	In summary we can group our classes on three major categories: CH wind (classes 0 and 1, colored in red), quiet or transitional wind (classes 2 and 3, colored in green), and transients (classes 4 and 5, colored in blue).
	
	\subsection{Verification of the DSOM classes}
	
\begin{figure}[h!]
	\begin{center}
		\includegraphics[width=18cm]{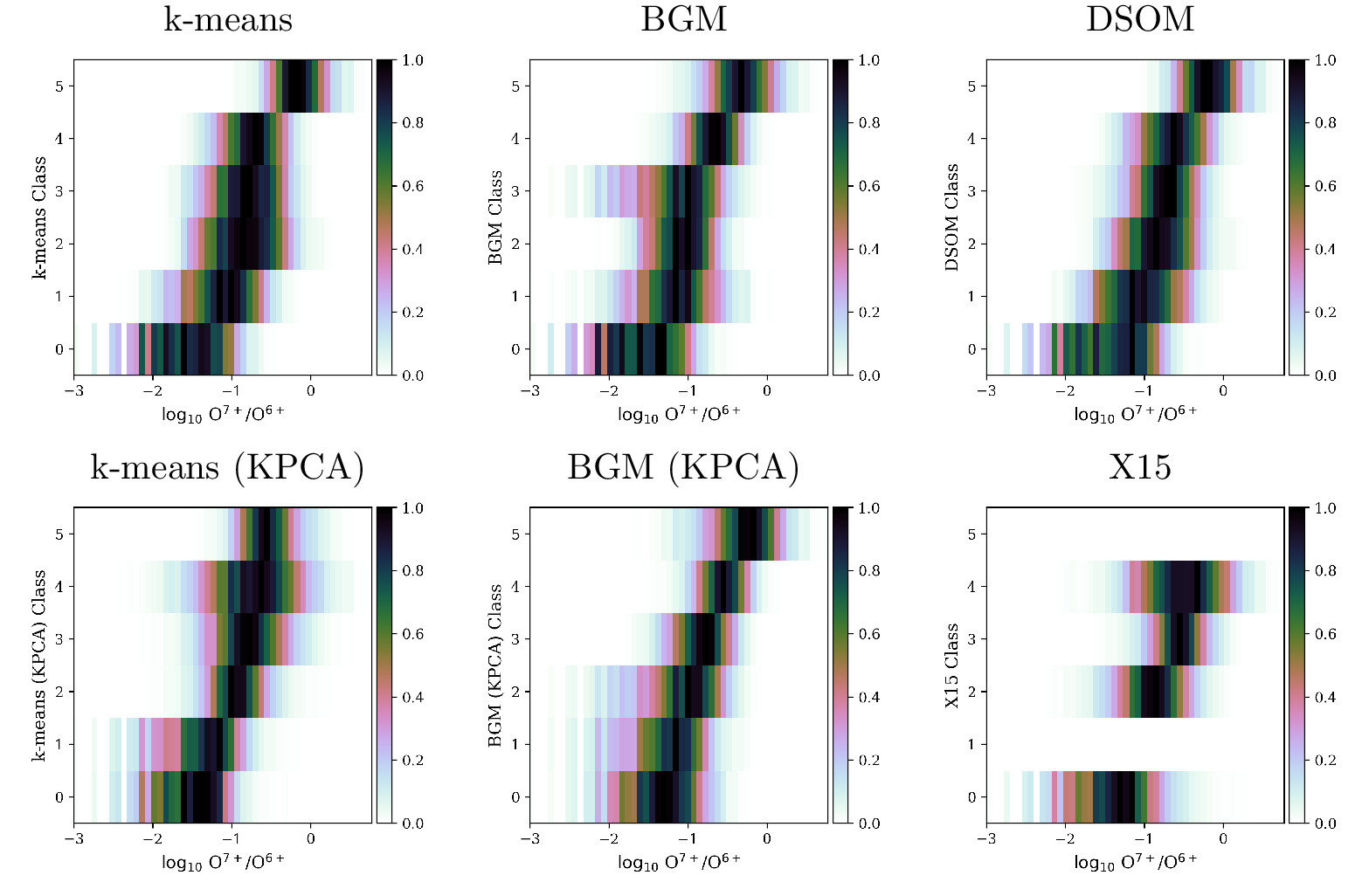}%
	\end{center}
	\caption{Histograms of the distribution of $\log_{10} O^{7+}/O^{6+}$ ratio on each one of the classes obtained by multiple classification methods and by the X15 classification.}\label{fig:Chist}
\end{figure}

\begin{figure}[h!]
	\begin{center}
		\includegraphics[width=18cm]{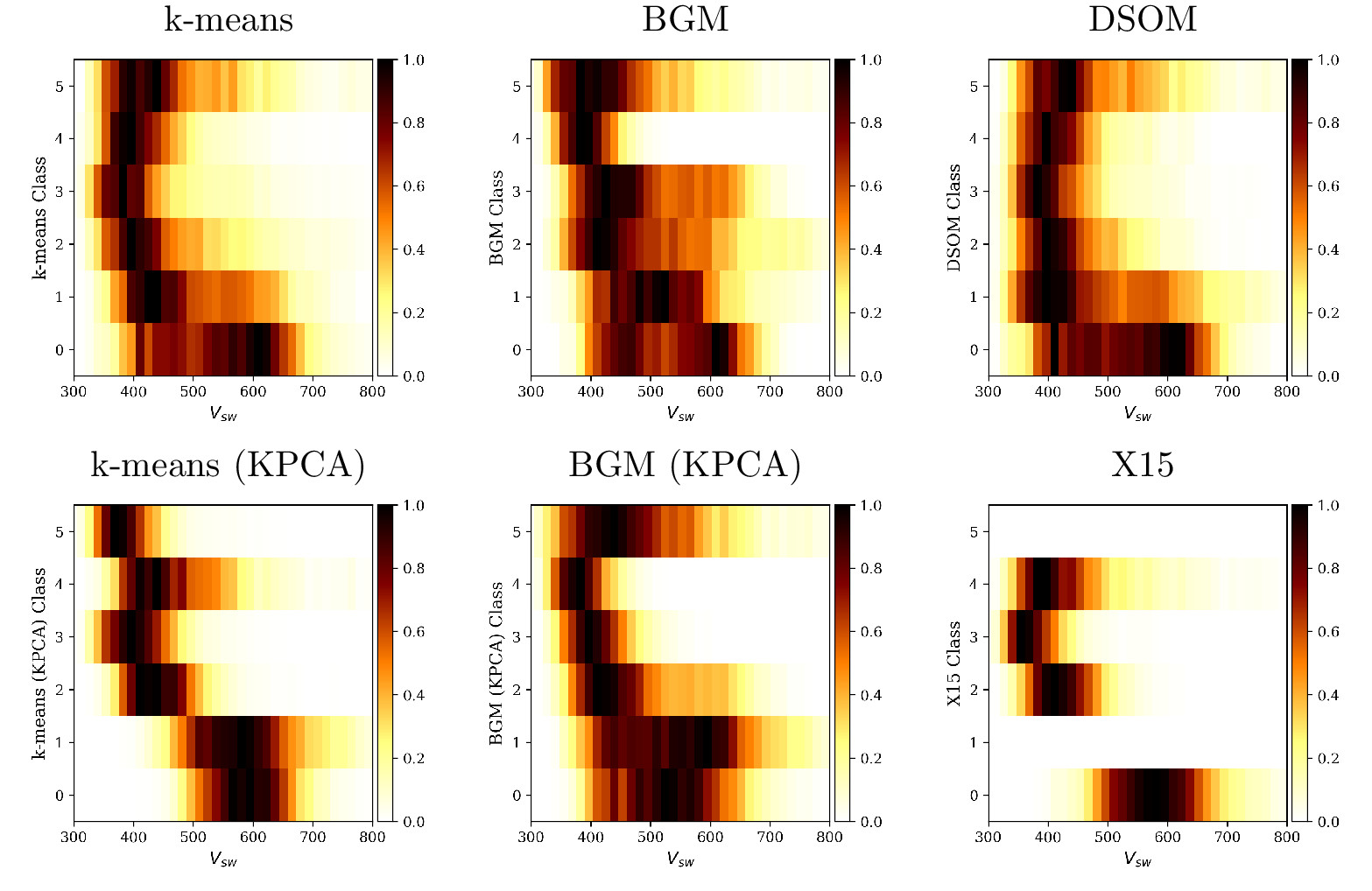}%
	\end{center}
	\caption{Histograms of the distribution of solar wind speed, $V_{sw}$, on each one of the classes obtained by multiple classification methods and by the X15 classification.}\label{fig:Vhist}
\end{figure}

	In addition to the interpretation of the maps presented in the previous section, we have extracted histograms of the occurrence frequency of $O^{7+}/O^{6+}$ ratio (Fig.\ref{fig:Chist}) and proton speed, $V_{sw}$ (Fig.\ref{fig:Vhist}). The panels in the figures contain the histograms for six (6) different categorizations: $k$-means (AE), $k$-means (KPCA), BGM (AE), BGM (KPCA), DSOM and the X15 classification. All the histograms have been normalized row by row (class by class), following the work done by \cite{Zhao2017}. This representation of the data is inspired by figure 5 of that paper, where the authors showed an important overlapping among different solar wind classes, and a bi-modal velocity distribution for coronal hole wind including an important population of slow wind.
	
	The assignment of class numbers by the clustering algorithms is random. We have sorted the classes so they present an ascending value of the $O^{7+}/O^{6+}$ ratio in Fig.\ref{fig:Chist}. It has been shown that solar wind originated in \textit{Coronal Holes} present very low values of the $O^{7+}/O^{6+}$, while at the other extreme transient events present very high $O^{7+}/O^{6+}$ ratios~\cite{Zhao2017, Zhao2009, Stakhiv2016}. Fig.\ref{fig:Chist} confirms the class identification we presented in the previous section.%

	\cite{VonSteiger2015,Bloch2020} examine the $O^{7+}/O^{6+}$ ratio in Ulysses data, which include abundant measures of wind originating from the polar CHs. Our data is composed of ACE observations from the ecliptic plane. For this reason, in all different classifications in Fig.\ref{fig:Chist}, including the X15 empirical categorization, class 0 does not reach $\log_{10} O^{7+}/O^{6+}\approx -2$, where the peak of points is observed in publications using Ulysses data. %

	In our data set, the majority of points can be mapped to \textit{Quiet Sun} (QS), conditions, i.e. slow solar wind.
	Even in these conditions, the DSOM method is able to sample and distribute enough data to each one of the classes. The BGM method applied to Kernel PCA transformed data also provide a good sample of the different classes, in particular for transient solar wind (classes 4 and 5). The X15 classification was designed with clear boundaries in $O^{7+}/O^{6+}$, for this reason the differences among the four classes is clear in the histograms. However, this observation contradicts the foot point back tracing performed by \cite{Zhao2017}: X15 shows almost no overlap in the distribution functions between the different classes, while the back tracing shows important overlaps. We express caution in the use of this classification to train any type of supervised machine learning technique, or in the evaluation of the accuracy of unsupervised techniques.
	
	Fig.\ref{fig:Vhist} shows how velocity is distributed among the different classes for each unsupervised classification method, and for the X15 categories. \cite{Zhao2017} remarks that the different classes are more difficult to identify using the solar wind speed histograms. We verify in these plots that three conditions are satisfied: 1) the classes we associate to the QS (class 2 and 3 in the DSOM classification) are associated to low velocity regions \cite{Neugebauer2002}, 2) high oxygen state ratios are associated with low solar wind speeds, and 3) CH wind has a highly spread velocity distribution, with two possible peaks around 400 km/s and 600 km/s \cite{Zhao2017}.
	
	The fact that class 0 and 1, that we associate to wind of CH origin, contains slow wind data points is particularly significant. \cite{DAmicis2015} has provided proof of the presence, at 1 AU, of highly Alfv\'{e}nic slow wind originating from the boundaries of coronal holes. This slow,  Alfv\'{e}nic wind  has the same composition signature and high cross helicity  that characterized the \textit{classic} fast Alfv\'{e}nic wind of CH origin, but presents lower speed and lower proton temperature. This way of visualizing our results seems to suggest that slow Alfv\'{e}nic wind is classified together with fast Alfv\'{e}nic wind in the classes that we associate to CH origin.
	
	Fig.\ref{fig:Vhist} shows that CH wind in the $k$-means and DSOM plots present a broad range of speeds, with a bimodal distribution. The BGM (KPCA) method separates these two populations in two different classes (0 and 1). the $k$-means (KPCA) method differentiates the fast solar wind, in the first two classes, from the slow wind in the remaining classes.
	
	Balancing the results from Fig.\ref{fig:Chist} and Fig.\ref{fig:Vhist} we conclude that the BGM (KPCA) and the DSOM are the techniques that approach the most the direct observations of the solar wind origins obtained by \cite{Zhao2017}. The X15 model creates a very sharp separation of solar wind types, with fast winds clearly segregated in class 0, slow winds in classes 1 and 2, and transients in class 3. The X15 model does not recognize that plasma of CH origin also contains an important population of slow winds.
	
	\subsection{Hit maps of empirical classifications}
	
\begin{figure}[h!]\centering
	\includegraphics[width=18cm]{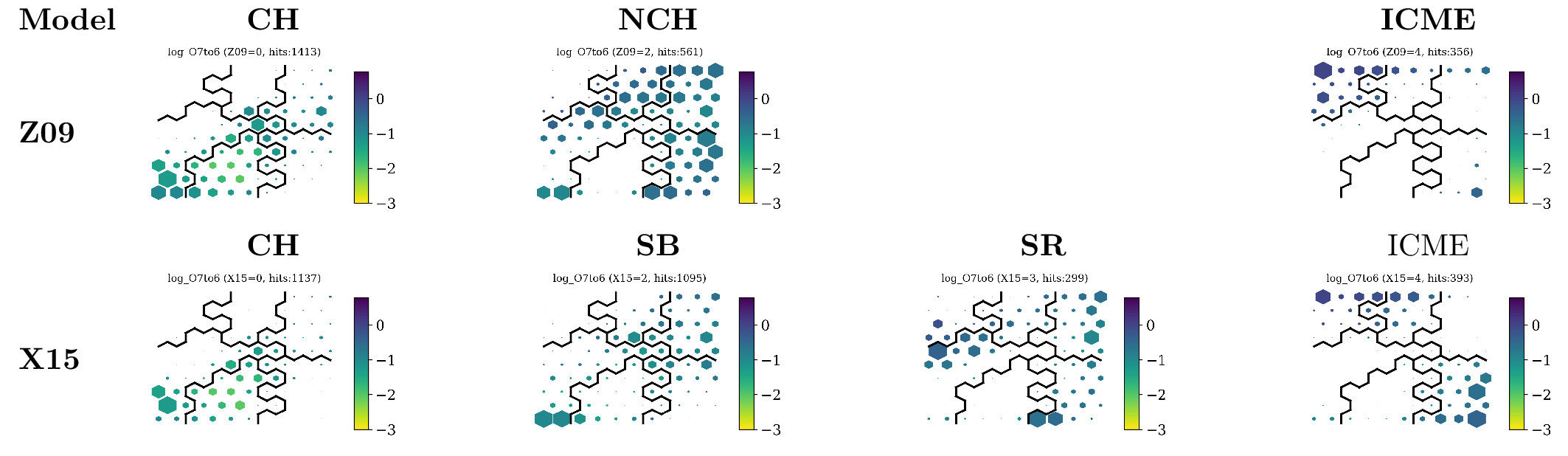}
	\caption{DSOM plots showing the activation of nodes for the different classes of the Z09 and X15 classifications. All maps are colored by the $\log_{10} O^{7+}/O^{6+}$ ratio, and the size of the hexagon represents the frequency of points, or number of hits.}\label{fig:SWZ09X15}
\end{figure}

	Another advantage of the SOM/DSOM method is that it can be used to visualize additional hidden statistics. Fig.\ref{fig:SWZ09X15} shows what nodes are activated by the Z09 and X15 classes. To perform this analysis, instead of using the full data set, we extract 3 subsets corresponding to the entries categorized as CH, ICME and NCH wind in the Z09, and CH, SB, SR and ICME in the X15 catalogs. Each one of these three (four) subsets is passed through the DSOM model and we observe how each one activates the nodes.
	
	We see that CH wind, in column 1 of the figure, activates very similar nodes for both classifications, in classes 0 and 1. Most of the hits are located on nodes where the absolute value of the cross-helicity $\sigma_c$ is the largest, i.e. in regions of open field lines associated with coronal holes.
	
	NCH wind from the Z09 classification is distributed over classes 2, 3 and 4, but also includes a node from class 1 characterized by an extremely negative cross-helicity. The same zone is activated by the SB class from X15. The two affected nodes also feature a very low $T_\text{exp}/T_p$. The X15 model splits solar wind points using hyperplanes in a three-dimensional space composed by $S_p$, $O^{7+}/O^{6+}$, and $T_\text{exp}/T_p$, None of those planes cuts the points in the $T_\text{exp}/T_p$ dimension \cite{Xu2015b}. However, in our maps this dimension seems to play an important role in the separation between quiet and CH winds.
	
	The X15 Sector Reversal (SR) class activates nodes at the boundaries of classes 1, 2 and 3. These nodes separate the quiet sun from the coronal hole wind, and coronal holes to transients. It also contains a large population of slow quiet solar wind.
	
	Finally transients, in both the Z09 and X15 categorizations, are associated to our class 4 and 5. However, a large portion of the X15 transients is associated to class 3 of the DSOM, particularly in nodes showing low proton temperatures and specific entropy $S_p$, characteristics of ICMEs.
	
	Fig.\ref{fig:SWZ09X15} shows also that, on average, the values of the $O^{7+}/O^{6+}$ ratio do not change radically among the nodes, except for small variations in the CH and the ICME classes of X15.
	
	\subsection{Quantitative comparison with empirical classifications}
	
	We have included a \textit{Matching Matrix} in Table \ref{tab:matchingmatrix} showing the frequency of occurrences of our model with respect to the Z09 and X15 classifications. Bold numbers in the table mark the highest common frequency and regular fonts mark the second highest frequency for each one of the columns. Matching matrices must not be confused with confusion matrices, as the later imply that there is a \textit{ground truth}.  Matching matrices are used in unsupervised learning to compare the frequency of occurrence of classes between models, so we can not perform additional metrics, like accuracy, precision, sensitivity or specificity.
	
	In this matrix we see that CH and SB categories from the X15 classification are mostly associated with classes 0, 1, 2 and 4 in the DSOM model, while TR winds are associated with classes 3 and 5. No particular class is clearly associated with SR winds, but the highest frequency is observed for class 0.
	
	CH in the Z09 classification are accurately associated with classes 0 and 1, but a big part of the NCH wind is also grouped in class 0. Transients are correctly distributed among classes 4 and 5 of the DSOM.
	
	We highlight that the X15 and Z09 models, the two classifications most used for the verification of machine learning results (see \cite{Camporeale2017b,Li2020}), are not fully compatible among themselves. A large number of CH winds from the Z09 classification is associated with SB winds in the X15 classification, and a considerable number of transients are cataloged as sector boundary crossings (SB).
	
	\subsection{Time series comparisons}
	
\begin{figure}[h!]
	\begin{center}
		\includegraphics[width=18.5cm]{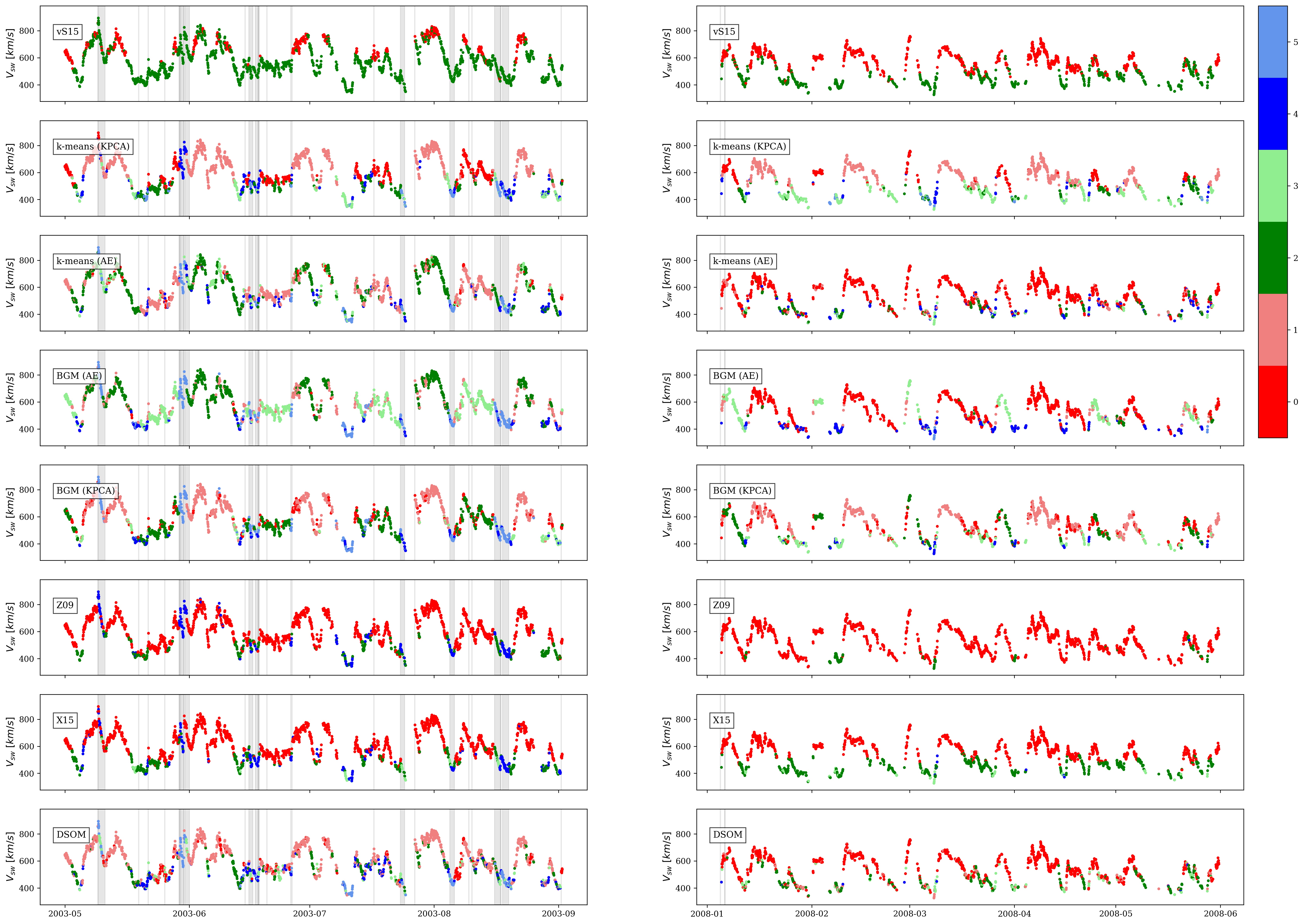}%
	\end{center}
	\caption{Solar wind speed observed by ACE in two windows of time: during high solar activity (left column) and during low solar activity (right column). Each row corresponds to a different solar wind classification method. Vertical gray zones and lines correspond to entries in the Richardson and Cane, UNH and CfA catalogs described in section \ref{sec:catalogs}.}\label{fig:timeseries}
\end{figure}

	A complementary method to compare the different classification techniques is to visually inspect windows of time and check, with the help of a human expert, that the time series are in agreement with the previous analysis. Fig.\ref{fig:timeseries} shows, in two columns, two windows of time of four months. The left column contains a high solar activity period, from May 2003 to September 2003, and the right column contains a period of low solar activity, between January 2008 and May 2008. Each one of the eight (8) rows contains a plot of the solar wind speed colored by a different classification method, from empirical models (Z09, vS15, and X15) to unsupervised methods ($k$-means, $k$-means, BGM, and DSOM). The colors of the empirical methods in the time series correspond to the labels assigned in Table \ref{tab:swtypes}, and the colors of the models were all assigned by manually ordering the classes following the frequency $\log_{10}(O^{7+}/O^{6+})$, from low values (low category number) to high values (high category number).
	
	In the same figure vertical gray zones correspond to Richardson and Cane ICME catalog entries \cite{Richardson2012}, and vertical lines to entries in the UNH and CfA catalogs.
	
	It is clear that among the empirical models, the vS15, based on observations by the Ulysses mission, is the most restrictive in the selection of CH origin winds, however during the plotted quiet time in the right column, which corresponds to the declining phase of the solar cycle, a significant part of the solar wind originates in coronal holes, and in fact High Speed Stream and Corotating Interaction regions, associated to wind of CH origin, are the main driver of geomagnetic activity during the declining phase of the cycle  \cite{tsurutani2006corotating, innocenti2011improved}. During both solar activity windows the Z09 and X15 models assign an important number of observations to coronal holes. \cite{VonSteiger2015} shows that the threshold used in the Z09 classification to identify coronal holes is not accurate and can misclassify NCH as CH. Both models accurately identify transients in the data. Quiet solar wind is more clearly visible during the low solar activity window in the X15 model.
	
	$k$-means (KPCA) and BGM (KPCA) correctly classify CH origin winds (classes 0 and 1). A clear transition between class 1, CH wind, and class 3 can be observed on both panels.  Transients are also well captured with classes 4 and 5. On the other hand, classifications based on the $k$-means (AE) and BGM (AE), do not show high accuracy in these two windows of time, but are able to detect transients. These two methods show difficulties in discerning QS winds from CH solar winds.
	
	The DSOM model shows good performances. The two classes associated with CH origin wind, classes 0 and 1, are more restrictive than the Z09 and X15 classes. Classes 4 and 5 distinguish between two different types of transients. ICMEs in these time windows are mainly associated with class 5, except for transients observed around 2003-05-20 and 2003-06-15. The model also detects a very slow transient around 2003-07-10. The 27 day solar period is also evident on the oscillations of the solar wind speed and the periodic nature of the solar wind types. In the low solar activity window the solar wind is more homogeneous and shows mainly CH and QS origin winds, as expected~\cite{tsurutani2006corotating}.
	
	Different classification methods lead to different classes with different properties. \cite{Roberts2020} performed detailed descriptions of the categorized solar wind classes based on the mean values observed in each subset of points. \cite{Zhao2017} shows that it is important to look at the frequency distribution and not only the mean. Our model shows that some features can present very large distributions inside a single class, even multiple peaks, as is the case of the solar wind speed for the CH classes.
	
	We will perform further refinements of the model and its interpretation in a future work. These preliminary results show the great potential of the techniques introduced in this paper. DSOMs show the variability of solar wind and how it can be visually characterized. The DSOM is a helpful guide in the study of the different types of solar wind, but is not necessarily an objective, unbiased and final classification method. In our current understanding, the main factor that determines classification results is the choice of the solar wind parameters used in the DSOM training. Choosing parameters that, according to previous studies and our physical understanding of the wind, can discriminate between specific wind types can guide the classification results. On the other hand, the possibility exists that an unsupervised classification methods such as the one used here will highlight the presence of solar wind types that could warrant future physical investigation. DSOMs open the possibility for a fast visual characterization of large and complex data sets.
	
	\section{Discussion}
	In this paper we show how the categorization of solar wind can be informed by classic unsupervised clustering methods and Self-Organizing Maps (SOM). We demonstrate that a single technique used in isolation can be misleading in the interpretation of automatic classifications. We show that it is important to examine the SOM lattices, in conjunction with solar wind composition and velocity distributions, and time series plots. Thanks to these tools we can differentiate classes associated with known heliospheric events.
	
	We are convinced that basic unsupervised clustering techniques will have difficulties in finding characteristic solar wind classes when they are applied to unprocessed data. A combination of feature engineering, non-linear transformations and SOM training leads to a more appropriate segmentation of the data points.
	
	The classification of the solar wind also depends on the objectives that want to be attained: if the goal is to classify the solar wind to study its origin on the Sun, features related to solar activity must be included in the model; however, if the goal is to identify geoeffectiveness, other parameters should be added to the list of features, including geomagnetic indices.
	
	In this work we have presented a first test of the capabilities of the SOMs for the analysis of data from a full solar cycle. Due to the extent of the work done, in this paper we introduce all the methods and techniques developed, but we leave for a future publication a more refined selection of all the model parameters, and the corresponding interpretation of the solar wind classification.
	
	Finally, we advocate for the creation of a catalog of foot point locations for every solar mission, that connect solar wind observations to points on the solar surface. Due to the uncertainty of on the exact foot point, such catalog should be composed of a set of probabilities for each possible solar origin. This ground truth will vastly improve the efficacy of our classification models, which in turn can be used to reduce the initial uncertainties of the catalog.
	
	All the tools and the techniques presented here can be applied to any other data set consisting of large amounts of points with a fixed number of properties. All the software and the data used in this work are freely available for reproduction and improvement of the results presented above.
	
	\section{Additional Requirements}
	
	\section*{Conflict of Interest Statement}

	The authors declare that the research was conducted in the absence of any commercial or financial relationships that could be construed as a potential conflict of interest.
	
	\section*{Author Contributions}
	
	JA created the software used in this work, built the data sets and wrote the manuscript. RD provided important insights into the use of the machine learning techniques, and performed revisions of the different drafts. MEI gathered information from external collaborators, provided insights into the data usage, and proofread different drafts. GL supervised the work. All authors contributed to manuscript revision, read and approved the submitted version.
	
	\section*{Funding}
	
	The work presented in this manuscript has received funding from the European Union’s Horizon 2020 research and innovation programme under grant agreement No 754304 (\href{https://www.deep-projects.eu}{DEEP-EST, www.deep-projects.eu}), and from the European Union’s Horizon 2020 research and innovation programme under grant agreement No 776262 (\href{https://www.aida-space.eu}{AIDA, www.aida-space.eu}).
	
	\section*{Acknowledgments}
	
	The authors would like to acknowledge the helpful advice and suggestions by Olga Panasenco (UCLA), Raffaella D'Amicis (INAF-IAPS), and Aaron Roberts (NASA-GSFC).
	
	We thank the teams of the ACE SWEPAM/SWICS/MAG/EPAM instruments and the ACE Science Center for providing the ACE data.
	
	\section*{Supplemental Data}
	
	All the software and the data used in this work can be found in the following git repository: \href{http://github.com/murci3lag0}{github.com/murci3lag0}.
	
	All original source code is distributed under MIT license.
	
	\section*{Data Availability Statement}
	\label{sec:repos}
	
	All processed data, including edited catalogs and enhanced data sets are available in our git repository: \href{http://github.com/murci3lag0}{github.com/murci3lag0}.
	
	All original data is distributed under Creative Commons license: CC BY 4.0.
	
	The original ACE data sets were downloaded from \href{ftp://mussel.srl.caltech.edu/pub/ace/level2/multi}{ftp://mussel.srl.caltech.edu/pub/ace/level2/multi}.
	
	A detailed description of all ACE multi-instrument data set entires can be found here: \href{http://www.srl.caltech.edu/cgi-bin/dib/rundibviewmultil2/ACE/ASC/DATA/level2/multi}{http://www.srl.caltech.edu/cgi-bin/dib/rundibviewmultil2/ACE/ASC/DATA/level2/multi}
	
\printbibliography

\section*{Tables}

\begin{table}\centering
	\begin{tabular}{@{}llrll@{}}
		\toprule
		\# & SW type & Condition & & Reference \\
		\midrule
		0 & CH & $\log_{10}{O^{7+}/O^{6+}}\leq$&$0.145$ & \cite{Zhao2009} \\
		2 & NCH & $0.145< \log_{10}{O^{7+}/O^{6+}} <$&$ 6.008e^{(-0.00578V_{sw})} $ & \\
		4 & TR & $\log_{10}{O^{7+}/O^{6+}}>$&$6.008e^{(-0.00578V_{sw})}$ & \\
		\midrule
		0 & CH & Not type TR, and & & \cite{Xu2015b}\\
		&  & $\log_{10}(S_p)>$&$ -0.525 \log_{10}(T_{\text{exp}}/T_p)$\\
		&  & &$- 0.676\log_{10}(V_A)$ \\
		&  & &$+ 1.74$ & \\
		2 & SB & Not type CH, TR, or SR & & \\
		3 & SR & Not type TR, and & & \\
		& & $\log_{10}(S_p)<$&$ -0.125 \log_{10}(T_{\text{exp}}/T_p)$ \\
		& & & $- 0.658\log_{10}(V_A)$ \\
		& & & $+ 1.04$ &  \\
		4 & TR & $\log_{10}(V_A)>$ & $ 0.055\log_{10}(T_{\text{exp}}/T_p)$ \\
		&  & & $+ 0.277 \log_{10}(S_p)$ \\
		&  & & $+ 1.83$ & \\ 
		\midrule
		0 & CH & $\log_{10}{O^{7+}/O^{6+}}\times\log_{10}{C^{6+}/C^{5+}}\leq$ & $0.01$ & \cite{VonSteiger2015}\\
		2 & NCH & Not type CH & & \\
		\midrule
		0 & CH & $1.25 \times \log_{10}{O^{7+}/O^{6+}}+6.75 <$& $\log_{10}(S_p)$ & \cite{Bloch2020}\\
		2 & NCH & Not type CH & & \\
		\bottomrule
	\end{tabular}
	\caption{Solar wind types and boundaries as defined by the empirical models: Z09, X15, vS15, and B20. The four types are: fast solar wind of coronal hole origin (CH), slow wind of non-coronal hole origin (NCH), transients, including ejecta, ICMEs, CIRs, MCs or other sudden jumps in solar wind parameters (TR), solar wind originated in the streamer belt (SB), and solar wind of sector reversal origins (SR). The ID value in the first column is arbitrary and has been chosen to simplify the visualization of our results.}
	\label{tab:swtypes}
\end{table}

\begin{table}\centering
	\begin{tabular}{@{}rlc@{}}
		\toprule
		ID & Name in the database  & AE+DSOM \\
		\midrule
		0 & proton\_speed & $\checkmark$ \\
		1 & proton\_density & $\checkmark$ \\
		2 & O7to6 & $\checkmark$ \\
		3 & C6to5 & $\checkmark$ \\
		4 & FetoO & $\checkmark$ \\
		5 & avqFe & $\checkmark$ \\
		6 & proton\_temp & $\checkmark$ \\
		7 & sigmac$^{(*)}$ & $\checkmark$ \\
		8 & Sp & $\checkmark$ \\
		9 & Va & $\checkmark$ \\
		10 & Tratio & $\checkmark$ \\
		11 & proton\_speed\_range & $\checkmark$ \\
		12 & Bn\_range & $\checkmark$ \\
		13 & FetoO\_range & $\checkmark$ \\
		14 & O7to6\_range & $\checkmark$ \\
		\midrule
		& Initial year & 1998 \\
		& Final year & 2011 \\
		& Neurons / encoding layer & [15, 10] \\
		& Optimizer & Adam \\
		& Learning rate & 0.042 \\
		\midrule
		& Lattice nodes &  10$\times$10 \\
		& $\epsilon$ &  0.005 \\
		& $\eta$ & 3.0 \\
		\bottomrule
	\end{tabular}
	\caption{List of features used for the AE+DSOM model. The logarithm of all quantities was used, except for the features marked with an asterisk (*). Bottom: data range and hyper-parameters of the AE and the SOM.}
	\label{tab:features}
\end{table}

\begin{table}\centering
	\begin{tabular}{|l|l|llllll|lll|}
		\cline{3-11}
		\multicolumn{2}{c|}{} & \multicolumn{6}{c|}{DSOM Class} & \multicolumn{3}{c|}{Z09} \\
		\cline{3-11}
		\multicolumn{2}{c|}{} &    0 &    1 &    2 &    3 &    4 &    5 &     CH &     NCH &     TR \\ \hline
		\multirow{4}{10pt}{\rotatebox[origin=c]{90}{X15}} & CH & \textbf{7727} & 3994 & 3330 &  \gray{125} &  502 &   47 & \textbf{14273} &   \gray{993} &   \gray{459} \\
		& SB & 7423 & \textbf{6295} & \textbf{7194} &  916 &  \textbf{950} &  \gray{138} & 12904 &  \textbf{9244} &   768 \\
		& SR & \gray{3233} &  \gray{392} & \gray{1404} &  \gray{584} &  \gray{434} &  157 &   \gray{877} &  4747 &   \gray{580} \\
		& TR & \gray{1263} &  \gray{575} &  \gray{387} & \textbf{2373} &  \gray{403} & \textbf{1362} &  \gray{1343} &  \gray{3011} &  \textbf{2009} \\
		\hline
		\multirow{4}{10pt}{\rotatebox[origin=c]{90}{DSOM Class}} & 0 &      &       &     &      &      &      & \textbf{13357} &  5966 & \gray{323} \\
		& 1 &      &       &     &      &      &      &  \textbf{9040} &  2213 &     \gray{3} \\
		& 2 &      &       &     &      &      &      &  \textbf{6173} &  5789 &   \gray{353} \\
		& 3 &      &       &     &      &      &      &   637 &  \textbf{2848} &   \gray{513} \\
		& 4 &      &       &     &      &      &      &   \gray{170} &   987 &  \textbf{1132} \\
		& 5 &      &       &     &      &      &      &    \gray{20} &   192 &  \textbf{1492}\\ \hline
	\end{tabular}
	\caption{Matching matrix comparing the DSOM, X15 and Z09 classifications. Values in bold (regular) font represent the highest (second highest) frequency for each column (row) in the top (bottom) half of the table.}
	\label{tab:matchingmatrix}
\end{table}

\section*{Figure captions}

\newcolumntype{C}{>{\centering\arraybackslash} m{4cm} }

\newcolumntype{C}{>{\centering\arraybackslash} m{4.5cm} }

\end{document}